\documentclass[longauth]{aa}
\usepackage{graphicx}
\usepackage{txfonts}
\usepackage{hyperref}
\usepackage{textcomp}
\usepackage{booktabs}
\usepackage{multirow}
\usepackage{color}
\usepackage{orcidlink}
\usepackage[switch]{lineno}
\usepackage[normalem]{ulem} 
\linenumbers
\usepackage{amsmath}

\newcommand{\hunit}{\ensuremath{\mathrm{km\,s^{-1}\,Mpc^{-1}}}}

\begin{document} 
\nolinenumbers

   \title{Study of the anisotropy of cosmic expansion on ZTF type Ia supernovae simulations}

   \author{Barjou-Delayre, C.
          \inst{1}\fnmsep\thanks{chloe.barjou-delayre@clermont.in2p3.fr}\orcidlink{0009-0000-8510-8982}
          \and
          Rosnet, P.\inst{1}\fnmsep\thanks{philippe.rosnet@clermont.in2p3.fr}\orcidlink{0000-0002-6099-7565}
          \and
          Ravoux, C.\inst{1}\orcidlink{0000-0002-3500-6635}
          \and
          Kebadian, R.\inst{2}\orcidlink{0009-0003-4093-1870}
          \and
          Aubert, M.\inst{1,3}\orcidlink{0009-0002-7667-8814}
          \and
          Ginolin, M.\inst{4}\orcidlink{0009-0004-5311-9301}
          \and
          Amenouche, M.\inst{5}\orcidlink{0009-0006-7454-3579}
          \and
          Bautista, J.\inst{2}\orcidlink{0000-0002-9885-3989}
          \and
          Burgaz, U. \inst{6}\orcidlink{0000-0003-0126-3999}
          \and
          Carreres, B.\inst{7}\orcidlink{0000-0002-7234-844X}
          \and 
          Castaneda Jaimes, J. \inst{8} \orcidlink{0000-0002-0987-3372}
          \and
          Dimitriadis, G. \inst{9}\orcidlink{0000-0001-9494-179X}
          \and
          Feinstein, F.\inst{2}\orcidlink{0000-0001-5548-3466}
          \and
          Fouchez, D.\inst{2}\orcidlink{0000-0002-7496-3796}
          \and
          Galbany, L. \inst{10,11}\orcidlink{0000-0002-1296-6887}
          \and
          Ganot,C.\inst{3}\orcidlink{0009-0000-6300-9174}
          \and
          Graham, M. \inst{12}\orcidlink{0000-0002-3168-0139}
          \and
          Groom, S.L. \inst{13}\orcidlink{0000-0001-5668-3507}
          \and
          Goobar, A. \inst{14}\orcidlink{0000-0002-4163-4996}
          \and
          Johansson, J.\inst{14}\orcidlink{0000-0001-5975-290X}
          \and
          Kasliwal, M.M. \inst{8}\orcidlink{0000-0002-5619-4938}
          \and
          Kim, Y.-L. \inst{15}\orcidlink{0000-0002-1031-0796}
          \and
          Müller-Bravo, T.E \inst{16,17}\orcidlink{0000-0003-3939-7167}
          \and
          Popovic, B. \inst{18} \orcidlink{0000-0002-8012-6978}
          \and
          Racine, B. \inst{2} \orcidlink{0000-0001-8861-3052}
          \and 
          Regnault, N. \inst{19}\orcidlink{0000-0001-7029-7901}
          \and
          Rehemtulla, N. \inst{20,21,22}\orcidlink{0000-0002-5683-2389}
          \and
          Rigault, M. \inst{3}\orcidlink{0000-0002-8121-2560}
          \and
          Riddle, R.L.\inst{23}\orcidlink{0000-0002-0387-370X}
          \and
          Sollerman, J.\inst{14}\orcidlink{0000-0003-1546-6615}
          \and
          Townsend, A. \inst{24}\orcidlink{0000-0001-6343-3362}
          \and 
          Trigui, A. \inst{3}\orcidlink{0009-0005-9981-8121}}

   \institute{Université Clermont Auvergne, CNRS/IN2P3, LPCA, F-63000 Clermont-Ferrand, France
         \and
         Aix Marseille Université, CNRS/IN2P3, CPPM, Marseille, France
         \and
         Univ Lyon, Univ Claude Bernard Lyon 1, CNRS, IP2I Lyon/IN2P3, UMR 5822, F-69622, Villeurbanne, France
         \and
         Institute of Astronomy and Kavli Institute for Cosmology, Madingley Road, Cambridge CB3 0HA, UK
         \and
         Centre Spatial de Liège, Université de Liège, Avenue du Pré-Aily, 4031 Angleur, Belgium
         \and 
         School of Physics, Trinity College Dublin, College Green, Dublin 2, Ireland
         \and 
         Department of Physics, Duke University Durham, NC 27708, USA
         \and
         Division of Physics, Mathematics and Astronomy, California Institute of Technology, 1200 E. California Blvd, Pasadena, CA 91125, USA
         \and
         Department of Physics, Lancaster University, Lancaster, LA1 4YB, UK
         \and
         Institute of Space Sciences (ICE-CSIC), Campus UAB, Carrer de Can Magrans, s/n, E-08193 Barcelona, Spain.
         \and
         Institut d'Estudis Espacials de Catalunya (IEEC), 08860 Castelldefels (Barcelona), Spain
         \and
         Division of Physics, Mathematics, and Astronomy, California Institute of Technology, Pasadena, CA 91125, USA
         \and 
         IPAC, California Institute of Technology, 1200 E. California Blvd, Pasadena, CA 91125, USA
         \and
         The Oskar Klein Centre, Department of Astronomy,AlbaNova, SE-106 91 Stockholm , Sweden
         \and
         Department of Astronomy \& Center for Galaxy Evolution Research, Yonsei University, Seoul 03722, Republic of Korea
         \and
         School of Physics, Trinity College Dublin, The University of Dublin, Dublin 2, Ireland
         \and 
         Instituto de Ciencias Exactas y Naturales (ICEN), Universidad Arturo Prat, Chile
         \and
         School of Physics and Astronomy, University of Southampton, Southampton SO17 1BJ, UK
         \and 
         Sorbonne Université, Université Paris Cité, CNRS, Laboratoire de Physique Nucléaire et de Hautes Energies, 4 Place Jussieu, 75252 Paris, France
         \and
         8 Department of Physics and Astronomy, Northwestern University, 2145 Sheridan Road, Evanston, IL 60208, USA
         \and
         Center for Interdisciplinary Exploration and Research in Astrophysics (CIERA), 1800 Sherman Ave., Evanston, IL 60201, USA
         \and
         NSF-Simons AI Institute for the Sky (SkAI), 172 E. Chestnut St., Chicago, IL 60611, USA
         \and
         Caltech Optical Observatories, California Institute of Technology,Pasadena, CA 91125, USA
         \and
         Institut f\"ur Physik, Humboldt-Universit\"at zu Berlin, Newtonstr. 15, 12489 Berlin, Germany
         }

   \date{Received; Accepted}

\titlerunning{Study of the anisotropy of the cosmic expansion SNe Ia on simulation}
\authorrunning{C. Barjou-Delayre et al.}

 
  \abstract
   {The cosmological principle assumes the isotropy of the Universe at large scales. It is a foundational assumption in the $\Lambda$CDM model, which is the current standard model of cosmology. Recent tensions justify investigating the possibility of anisotropies in the Universe.}
   {The large sky coverage achieved by the Zwicky Transient Facility survey (ZTF) allows us to test the veracity of the cosmological principle by using observations of Type Ia supernovae (SNe Ia). In this article, we develop a methodology to measure potential anisotropies in the Hubble constant $H_0$. We test our method on realistic simulations of the second data release (DR2) of ZTF-SNe Ia in which we introduced a dipole.}
   {We develop a method to introduce a dipole in the simulations as well as to recover it. We test a potential $H_0$-dependency of our method and vary the amplitude of the dipole. We analyse the impact of introducing the large-scale structures in the simulations and the efficiency of using a volume-limited sample, which is an unbiased subset of the ZTF-SN Ia sample. We finally build an error model applied to the recovered dipole amplitude ($\Delta H_0$) and its direction ($\alpha_0$, $\delta_0$).}
   {Based on the various systematic studies carried out, our analysis recovers an input dipole of an amplitude of $3~\hunit$ with an error on the amplitude of $0.33~\hunit$, and $3.4^\circ$ and $6.1^\circ$ for the right ascension and declination. A null-fit amplitude test with simulation including large-scale structures shows that any dipole amplitude measured below an amplitude of $1.06~\hunit$ on real data would be compatible at 95\% with $\Lambda$CDM due to the sample variance of peculiar velocities. Moreover, the resulting dipole is independent of the chosen $H_0$ value and sky coverage. This paper paves the way for a future precise ZTF dipole investigation. }
   {}


   \maketitle
%

\section{Introduction}
\label{sec:intro}
The Lambda cold dark matter model ($\Lambda$CDM) is currently the most widely accepted description of the Universe, in  agreement with numerous observations. However, some tensions have emerged in the last decade, and certain assumptions can be questioned \citep{Abdalla_2022,Kumar_Aluri_2023}. The tension on the Hubble constant $H_0$ is the most striking, reaching a $5\sigma$ difference between the early and late Universe measurements (\cite{DiValentino_2021}). The late Universe measurements are performed using type Ia supernovae (SNe Ia) after calibration of their intrinsic luminosity with the period-luminosity relation of Cepheids. This measurement gives a value of $H_0^{\text{SHOES}}=73.04 \pm 1.04~\hunit$ \citep{Riess_2022}. The early Universe measurements are made using cosmic microwave background (CMB) observations, which are sensitive to the Hubble parameter $H$ at the time of recombination. It is then propagated to our current expansion rate $H_0$ using $\Lambda$CDM. For example, \cite{Planck_2020} measures: $H_0^{\text{Planck}}=67.66 \pm 0.42~\hunit$. The difference between these two measurements could be explained by an alternative cosmological model treating differently the early and late Universe. Alternatively, it can arise from potential systematic errors in the SNe Ia measurement, such as a correlation between the SNe Ia environment and their luminosity \citep{Rigault_2015}. Over the years, other measurements with various methods have been performed (Tip of the Red Giant Branch; \cite{Freedman_2025}, gravitational lensing of quasars; \cite{Chen_2019}, Baryon Acoustic Oscillation (BAO) measurements; \cite{DESI_II_2025}...). The precision of those measurements keeps the debate open, and no known systematic error can explain this tension. More recently, the BAO measurement from the Dark Energy Spectroscopic Instrument (DESI) \citep{DESI_2025} showed a hint for evolving dark energy, weakening even more the $\Lambda$CDM model.

The anomalies of the $\Lambda$CDM model \citep{Peebles_2025} justify reconsidering the main assumptions of this model. Among them, the cosmological principle assumes that the Universe is isotropic and homogeneous at large scales. Conversely, the CMB measurements from COBE \citep{COBE} and Planck \citep{Planck_2020} motivate an isotropy violation. This dipole seen in the CMB is generally explained by the motion of the Local Group (LG) relative to the CMB, due to the gravitational attraction of a nearby over-density \citep{Nusser_2011}. However, the various analyses addressing this find a significant dipole amplitude, which cannot be explained only by the motion of the LG \citep{Bolejko_2016}. To explain these measurements, some exotic models have emerged, such as an anisotropic Universe. The Bianchi-I model is the best-known and predicts a quadrupole in the expansion. This model is used in \cite{Verma_2024}, where the authors found slight evidence for a cosmic preferential axis. More complex inhomogeneous models, e.g., the Szekeres~\cite{Celerier_2024}, predict a time and spatial evolution of the Hubble parameter.

Computing the values of $H_0$ in different directions is a model-independent way of testing anisotropies of the cosmic expansion. With this method \cite{Boubel_2025} showed a 3\% variation of $H_0$ using the galaxy catalog Cosmicflows-4 \cite{Cosmicflow_2023}. However, they could not distinguish it from a possible bulk flow. \cite{Stiskalek_2025} measured a $4.1\pm 0.9\%$ $H_0$ dipole with the Cosmicflows-4 sample and $2.3\pm0.6\%$ with the Pantheon+ sample \cite{Pantheon_2022}. They note that this dipole is likely due to local flow features (or possibly systematic), but also that replacing the dipole model with a bulk flow model is consistent with the data and the $\Lambda$CDM model. However, the different analyses do not converge on the same dipole direction or amplitude and the question of potential expansion anisotropy is still under debate.

SNe Ia are well suited to such an analysis because their intrinsic luminosity makes them standardizable candles, enabling the determination of $H_0$, or the deceleration parameter $q_0$, whose anisotropy was studied in \cite{Cowell_2023}. Studying anisotropies requires a homogeneous survey and a large footprint to conduct analyses in different parts of the Universe while minimizing sky coverage biases. Previous studies have generally been performed on combinations of different SN Ia samples, such as Pantheon$+$ \citep{Scolnic_2022}, which uses 18 different surveys to form a sample of 1701 SNe Ia in the redshift range $0.01 < z < 2.26$. In our analysis, we mimic the ZTF survey, which gives a homogeneous sampling of SNe Ia in the Northern sky. This survey, and more particularly the second data release \citep{Rigault_2025_ov}, contains $3,628$ SNe Ia observed with the same instrument. Using a unique survey is a significant advantage in avoiding calibration issues and ensuring better control of systematic biases. The unique sample of ZTF allows to compute the value of $H_0$ in different sky directions. The goal of our analysis is to develop a methodology to fit possible $H_0$ anisotropies on ZTF realistic simulations, a method that will be applied later on real data.

This article is organized as follows: we present in Sec.~\ref{sec:data_stand} the ZTF second SN Ia data release (ZTF SN Ia DR2) and the associated realistic simulation we use in this work. In Sec.~\ref{sec:standardization}, we detail the SNe Ia standardization process. Sec.~\ref{sec:H_0_study} gives our method to introduce and fit an $H_0$ dipole. In Sec.~\ref{sec:systematic}, we test for potential systematic effects and parameter variations that could affect our analyses. Finally, in Sec.~\ref{sec:discussion} and Sec.~\ref{sec:conclusion} we interpret our results and draw the conclusions of our work.

\section{ZTF SNe Ia simulations and standardization}
\label{sec:data_stand} 

This analysis is based on simulation of the ZTF SN Ia DR2. In this section, we describe the ZTF SN Ia DR2 sample and the simulations of the survey using \texttt{skysurvey}. Finally, we explain the standardization of the SNe Ia to use them for a cosmological analysis.

\subsection{The second ZTF SN Ia data release}
\label{sec:ZTF_DR2}

The ZTF survey is a survey covering the Northern sky above a declination of $-30^\circ$ \citep{Bellm_2019}, using the Samuel Oschin 48-inch Schmidt Telescope based at Palomar Observatory in California (\cite{Graham_2019,Dekany_2020,Masci_2019}). ZTF has a large field of view of 47 deg$^2$, a camera made of 16 CCDs, and uses three filters in the $g$, $r$ and $i$-band. ZTF can observe the northern sky in 2 to 3 days. This high cadence provides an excellent light-curve sampling and a typical first detection two weeks before maximum light of an SN Ia, as presented in \cite{Rigault_2025_lc}.

In this paper, we simulate the ZTF SN Ia DR2 sample, which compiles SNe Ia detected between March 2018 and December 2020. It is composed of 3628 spectroscopically confirmed SNe Ia, mainly at redshift $z <0.15$. The Bright Transient Survey (BTS, \cite{BTS_I,BTS_II}) complements ZTF with a spectroscopic follow-up. It classifies transient objects and applies quality cuts on the number of observations around peak light and sky visibility, as detailed in \cite{BTS_II}. They obtain completeness fractions estimated at $97\%$, $93\%$, and $75\%$ for objects brighter than 18 mag, 18.5 mag, and 19 mag, in the three ZTF bands. Most of the ZTF SN Ia DR2 objects are spectroscopically typed with the SED machine (\cite{Rigault_2019,Kim_2022,Blagorodnova_2018}). Some spectra are collected from other telescopes, such as LT or P200. The ZTF SN Ia DR2 is described in more detail in \cite{Rigault_2025_ov}. As pointed out by \cite{Lacroix_2025}, the calibration quality of this sample is insufficient to perform a cosmological analysis. The next data release, named ZTF SN Ia DR2.5, will contain an additional calibration effort. Before applying it to real data, we test our method on realistic simulations of the ZTF SN Ia DR2.5. In a future paper, we will apply this analysis to the ZTF SN Ia DR2.5 sample. In our simulation, we expect that current calibration challenges have been solved and do not try to model the effects described in \cite{Lacroix_2025}. 

\subsection{Simulations}
\label{sec:simulations}
To simulate the ZTF SN Ia DR2.5 sample, we use the python package \texttt{skysurvey}\footnote{\url{https://github.com/MickaelRigault/skysurvey}} which generates astronomical targets as a survey would observe them \citep{Amenouche_2025}. To mimic observed SN Ia light-curves, we use the SALT2 SN Ia model \citep{Guy_2007,guy_2010} in its version 2.4 \citep{Betoule_2014}. The simulation software also accounts for ZTF observing conditions, detailed in \cite{Amenouche_2025}. We produce 20 independent simulations of the ZTF SN Ia DR2.5 data set by using the SN Ia rate measurement of ZTF, equal to $2.35 \times 10^{-4}~\text{Gpc}^{-3}~\text{yr}^{-1}$~\citep{BTS_II}. 
For each simulated SN Ia, its stretch, $x_1$, which corresponds to the width of the light-curve, is generated from a bimodal Gaussian distribution following \cite{Ginolin_2025_stretch}. The SN Ia color $c$, defined as the difference in magnitude between B and V bands \citep{Phillips_1993,Tripp_1998} follows a Gaussian with a distribution tail \citep{Ginolin_2025_color,Scolnic_2016}.

We generate approximately the same number of SNe Ia as in the ZTF SN Ia DR2.5 sample. This is performed after applying the same cuts as for observations, detailed in \cite{Amenouche_2025}: $x_1 \in [-3, 3]$, $c \in [-0.2, 0.8]$, $\sigma_{x_1}<1$, $\sigma_{c}<0.1$, $\sigma_{t_0}<1$, where $t_0$ is the peak-luminosity date in MJD, and a fit probability greater than $10^{-7}$. Once generated by \texttt{skysurvey}, the light-curves are fitted with the \texttt{sncosmo}\footnote{\url{https://github.com/sncosmo/sncosmo}} library. To reproduce data quality cuts, we keep only SNe Ia with at least seven detections, including two light-curve pre-peak and two post-peak detections, and we require detection in two different filters within the phase range $\phi \in [-10,40]$, where $\phi$ is the time difference in days between a light-curve point and the peak position ($t_0$). We reject all the detections with a signal-to-noise ratio lower than 5, as for the ZTF SN Ia DR2.5, using the same error on the flux presented in \cite{Amenouche_2025}:
\begin{equation}
    \sigma_f=\sqrt{s_{b}^{2} + \frac{|f|}{\mathrm{gain}} + (f \times \sigma_{\rm calib})^{2}}
\end{equation}
where $s_b$ is the sky brightness, $f$ the simulated flux, gain is the CCD gain equal to $\approx 6.2$, and $\sigma_{\rm calib}$ the calibration precision of ZTF SN Ia DR2.5 equal to $0.025,0.035$ and $0.06$ for the $g$, $r$ and $i$ bands respectively. The observing logs are used to reproduce the cadence of observation. These logs contain the field of observation, the limiting magnitude, and any possible technical issues. 

At first for the analyses detailed in Sec.~\ref{sec:Dipole_analyze} we used these 20 simulations which do not include the large-scale structure. In a second place, the \texttt{skysurvey} software can also generate SNe Ia following an input halo catalog. In that case, the effect of large-scale structure evolution is included, and each SN Ia has an associated peculiar velocity. With this method, we create 27 simulations of ZTF SN IA DR2.5, which include large-scale structures, using a halo catalog from the Uchuu N-body simulation~\citep{Ishiyama2020}. This simulation has a box size of $2~\text{h}^{-1}\text{Gpc}$ and is cut into 27 sub-grids to reproduce the volume of the ZTF survey, as done in \cite{Carreres_2023}. This reduction gives a maximum redshift of $0.11$ for each SN Ia simulation. To cover the $z>0.11$ range, the SNe Ia radial peculiar velocities are drawn randomly from a Gaussian distribution with a $250~\text{km}\text{s}^{-1}$ dispersion. In our study, we aim to see the impact of the SNe Ia peculiar velocities. We do not correct the peculiar velocities as in~\cite{Carreres_2025}; we leave this analysis for a future paper. The simulations that include large-scale structures are used only in Sec.~\ref{sec:large_scale_strucutre}.

\subsection{Cosmological distances from SNe Ia}
\label{sec:standardization}

SNe Ia are not perfect standard candles; thus, they need to be standardized to be used for accurate cosmology. The standardization process is performed using the SALT2 parameters presented in Sec.~\ref{sec:ZTF_DR2}: $x_1$, $c$, and $x_0$, the overall light-curve amplitude. We use the widely-known \citeauthor{Tripp_1998} relation:
\begin{equation}
\label{eq:standardization}
      \mu_{\rm obs} = m_B -M_B+\alpha x_1 - \beta c,
\end{equation}
where $m_B=-2.5\text{log}_{10}(x_0)+10.5$, the constant 10.5 is determined by \texttt{sncosmo} for the SALT2.4 model, $\beta$ is the color coefficient, $\alpha$ the stretch coefficient, and $M_B$ is the absolute SN Ia magnitude in the $B$-band. This standardization aims to correct the SNe Ia diversity to use them as standard candles. To do so, we fit the standardization parameters as follows: $M_B$, $\alpha$, $\beta$ from the observed distance modulus $\mu_{\rm obs}$. Compared to \cite{Ginolin_2025_stretch}, we do not use the $\gamma p$ term from SN Ia environmental dependencies because we do not include the SN Ia environment in our simulations. For cosmological analysis, we compute the distance modulus of a SN Ia from the luminosity distance $d_l$ following:
\begin{equation}
\label{eq:mutheo}
      \mu_{\rm cosmo} = 5 \text{log}(d_l(\text{Mpc}))+25 ;
\end{equation}
In the context of a flat $\Lambda$CDM model: 
\begin{equation}
\label{eq:dl}
      d_l=\frac{\text{c}}{H_0}(1+z)\int^{z}_0\frac{d\zeta}{\sqrt{((1+\zeta)^{3}-1)\Omega_{\rm m}+1}},
\end{equation}
where c is the speed of light, $\Omega_{\rm m}$ is the matter density parameter. For $H_0$ and $\Omega_{\rm m}$ we used the \textit{Planck}18 values \citep{Planck_2020}: $H_0=67.66~\hunit$ and $\Omega_{\rm m}=0.30966$. After the SN Ia light-curve fit, we fit the standardization coefficients $\beta$, $\alpha$, $M_B$ by using the following log-likelihood:
\begin{equation}
\label{eq:L_stand}
      \mathcal{L}(\alpha,\beta,M_B,\sigma_{\rm int})=-\frac{1}{2}\sum_{i=1}^{N_{\text{SNIa}}}\left(\left(\frac{\mu_{i,{\rm obs}}-\mu_{\rm cosmo}(z_i)}{\sigma_i}\right)^{2}+\text{log}(2\pi \sigma_{i}^{2})\right),
\end{equation}
with $N_{\text{SNIa}}$ the number of SNe Ia in the simulation survey and $\sigma_i$ the distance modulus uncertainty given by:
\begin{equation}
    \sigma_i= \sqrt{\sigma_{\mathrm{SALT},i}^{2}+\left(\frac{5 \sigma_{z_i}}{(1+z_i)\text{log}(10)}\right)^{2}+\sigma_{\rm int}^{2} },
\end{equation}
with $\sigma_{\mathrm{SALT},i}$ the error of the SALT2 fit:
\begin{align}\label{eq:sig_salt}
    \sigma_{\mathrm{SALT},i}^{2}= &\sigma_{m_B,i}^{2}+\beta^{2}\sigma_{c,i}^{2}+\alpha^{2}\sigma_{x_1,i}^{2}-2\beta V_i(m_B,c)\\
    &-2\beta \alpha V_i(c,x_1)+2\alpha \beta V_i(x_1,m_B) \nonumber,
\end{align}
where $\sigma_{m_B,i}$, $\sigma_{c,i}$, $\sigma_{x_1,i}$ are the errors of the SALT parameters. $V_i$ is the covariance matrix provided by SALT, here $\sigma_{z_i}=5 \times 10^{-4}$ is the uncertainty on SN Ia redshift. We do not account for the peculiar velocity uncertainties even when we introduce the large-scale structures in the simulations. Finally, $\sigma_{\rm int}$ is the intrinsic scatter of the SN Ia magnitude which account for unknown dispersion effects. This parameter is fitted along with $\alpha$, $\beta$ and $M_B$ in the likelihood of Eq.~\ref{eq:L_stand}.


\section{Study of a dipole anisotropy of the Hubble constant}
\label{sec:H_0_study}

\subsection{Methodology to introduce a dipole}
\label{sec:methodo_dip}

This section explains how we use a simulated ZTF sample, without the large-scale structure integrated in the simulations, to study a possible anisotropy in $H_0$. We test three methods to introduce a dipole in $H_0$:  at the level of the luminosity distances $d_l$, the redshift $z$, and the apparent magnitude $m_B$.

\subsubsection{Reference method for introducing a dipole: $d_l$-method}
\label{sec:method_dl}

The most rigorous way to introduce a dipole effect is through the luminosity distance $d_l$ according to:
\begin{equation}
\label{eq:dl_method}
      d_l'=\frac{c(1+z)}{H_{0}'}\int^{z}_0\frac{d\zeta}{\sqrt{((1+\zeta)^{3}-1)\Omega_{\rm m}+1}},
\end{equation}
with $H_0'=H_0+\Delta H_0 \cos(\Delta \theta)$, where $\cos(\Delta \theta)=\sin(\delta_0)\sin(\delta_i)+\cos(\delta_0)\cos(\delta_i)\cos(\alpha_0-\alpha_i)$ with $\alpha_i$, $\delta_i$ the equatorial position (RA, DEC) of the SN Ia and $\alpha_0$, $\delta_0$ the position of the dipole direction, and $\Delta H_0$ is the amplitude of the dipole. The value of the Hubble constant $H_0$ is the same as the one in Sec.~\ref{sec:standardization}. This method does not rely on any specific assumption but is computationally expensive. Indeed, because the SN Ia luminosity distance is set during the simulation stage, a new simulation is required for each dipole direction.

\subsubsection{Alternative method for inserting a dipole}
To overcome the computational issue of the $d_l$-method, we propose two alternative methods based on propagating $H_0$ modulations at the redshift level for one method and to at the magnitude level for the other one. 
\paragraph{$z$-method}
\label{sec:methodo_dip_z}

The first method is based on redshift modification after the simulations of SN Ia light-curves:
\begin{align}
\label{eq:dip_z}
      \text{c}z&=H_0d\\
      \text{c}z'&=H_0'd =(H_0+\Delta H_0 \cos(\Delta \theta))\;d, \\
      z'&= \left(1+\frac{\Delta H_0 \cos(\Delta \theta)}{H_0}\right)\;z,
\end{align}
with $d$ the comoving distance, c the speed of light and $z$ the initial redshift of the SN Ia, with $\cos(\Delta \theta)$ as defined in Sec.~\ref{sec:method_dl}. This method is only valid for very low redshifts as it uses the simplified Hubble relation. However, it is still used in literature. Here, we aim to prove that this method gives biased results even for a low-redshift survey.

\paragraph{$m_B$-method}
\label{sec:methodo_dip_mb}

As presented for the $z$-method, the $m_B$-method is not too computationally expensive, unlike the $d_l$-method. The $m_B$-method modifies the simulated SN Ia magnitude in the $B$-band, $m_B$, after the simulation of the SNe Ia, using: 

\begin{align}
    \mu=5\;\text{log}(d_l)+25=m_B-M_B,\\
    \mu'=5\;\text{log}(d_l')+25=m_B'-M_B,
\end{align}
\begin{align}
      d_l'=\frac{c(1+z)}{H_0'}\int^{z}_0\frac{d\zeta}{\sqrt{((1+\zeta)^{3}-1)\Omega_{\rm m}+1}}=\frac{H_0}{H_0'}d_l,\\
      \mu'=5\;\log(d_l)+25+5\;\text{log}\left(\frac{H_0}{H_0'}\right)=\mu+5\;\text{log}\left(\frac{H_0}{H_0'}\right).
\end{align}

By making the difference between $\mu$ and $\mu'$, we finally obtain: 
\begin{equation}
\label{eq:dip_mb}
      m_B'=m_B+5\;\text{log}\left(\frac{H_0}{H_0'}\right),
\end{equation}
with $H_0'=H_0+\Delta H_0 \cos(\Delta \theta)$.

Unlike the $d_l$-method, in these two methods the dipole is introduced after the light-curves have been generated, by modifying the parameters generated by the simulations. We produce a set of 20 simulations of the ZTF footprint, where we apply the $z$-method and $m_B$-method, and another set of 20 simulations where we modify the $d_l$-method at the generation of the SNe Ia also with the ZTF footprint. Our choice of dipole amplitude, for the three method, is based on taking a midpoint between the early and late Universe measurements of $H_0$, which is around $70~\hunit$. The amplitude is defined as the deviation of these central values from both the early and the late measurements, which correspond to $\Delta H_0 = 3~\hunit$.

\subsection{Dipole fitting}
\label{sec:Dipole_analyze}
We fit the dipole introduced on 20 independent simulations for the three methods introduced in the previous section: the $z$-method, $m_B$-method and $d_l$-method. Our goal is to determine the dipole amplitude ($\Delta H_0$) and its direction  ($\alpha_0$, $\delta_0$), by minimizing the following likelihood, which will be the same for all three methods of introducing a dipole :
\begin{equation}
\label{eq:L_dip}
      \mathcal{L}(\Theta)=-\frac{1}{2}\sum_{i=1}^{N_{\text{SNIa}}}\left(\frac{\mu_{i,\rm{obs}}(\Theta)-\mu_{\rm cosmo}(\Theta)}{\sigma_i}\right)^{2}+\text{log}(2\pi \sigma_{i}^{2}),
\end{equation}
with $\Theta={\alpha,\beta,M_B,\sigma_{\rm int},\Delta H_0,\alpha_0,\delta_0}$ the free parameters to fit. The fitted amplitude $\Delta H_0$ modifies the luminosity distance of each SN Ia by replacing $H_0$ in Eq.~\ref{eq:dl} by:
\begin{equation}
\label{eq:H0_prim}
     H_0'=H_0+\Delta H_0 \cos(\Delta \theta),
\end{equation}
with $H_0$ the same value mentioned in Sec.~\ref{sec:standardization} and $\cos(\Delta \theta)$ the same as in Sec.~\ref{sec:methodo_dip_z}.\\
The analysis is done following these steps, which ensure a good stability for the fitted dipole:
\begin{enumerate}
    \item We first fit the standardization parameters ($M_B$, $\alpha$, $\beta$, $\sigma_{\rm int}$) from Eq.~\ref{eq:standardization}.
    \item We performed a first dipole fit where our only free parameter is the amplitude of the dipole, and we test 12 different positions of the dipole, determined by splitting the sky with a Healpix schema\footnote{\url{https://healpix.sourceforge.io/documentation.php}} (nside=1). This first fit with 12 different dipole positions allows to determine a preferential area on the sky for a possible dipole. The dipole amplitude and location with the lowest $\chi^{2}$ value are selected and used as initial parameters for the next fit. This method gives a first estimate of the dipole amplitude and is necessary for fitting convergence.
    \item In a second fit, our free parameters are the dipole parameters (the amplitude $\Delta H_0$ and the coordinate $\alpha_0$, $\delta_0$) and the standardization parameters $M_B$, $\alpha$, $\beta$. We used as a fit first guess the values of $\alpha_0$, $\delta_0$, and $\Delta H_0$ determined from the previous step. The nuisance parameter $M_B$ is degenerate with $H_0$, which implies a correlation with dipole amplitude as detailed in Appendix.~\ref{appex:correlation}. For this reason, we have to fit the standardization parameters at the same time as the dipole parameters.
    \item The third fit consists in fitting the $\sigma_{\rm int}$ only, with the values of $M_B$, $\alpha$, $\beta$ and the dipole fixed from previous fit.  We have to fit $\sigma_{\rm int}$ independently, because fitting it with $M_B$, $\alpha$, $\beta$ introduces a bias in their values as shown in \citep{Kowalski_2008}.
\end{enumerate}

This methodology provides a data-based approach that allows probing all directions in the sky without any assumptions. As we focus only on comparing the three method in this section, we use the $x_1$, $c$ and $x_0$ generated from the simulations.

\subsection{Dipole fit results}
\label{sec:dip_result}

\begin{figure}
   \centering
   \includegraphics[width=0.9\columnwidth]{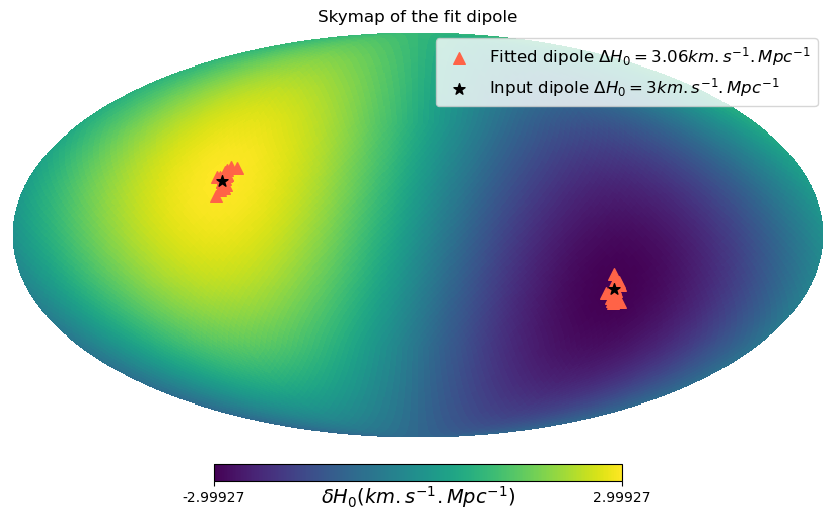}
      \caption{Skymap colored following the amplitude of the dipole, with the black stars representing the initial dipole direction of an amplitude of $3~\hunit$. The orange triangles are the fit directions for the 20 simulations when the $m_B$-method is used. The fitted amplitude in the legend corresponds to the median value over the 20 simulations.}
    \label{fig:skymap_mb}
\end{figure}

\begin{figure*}
   \centering
   \includegraphics[width=1.5\columnwidth]{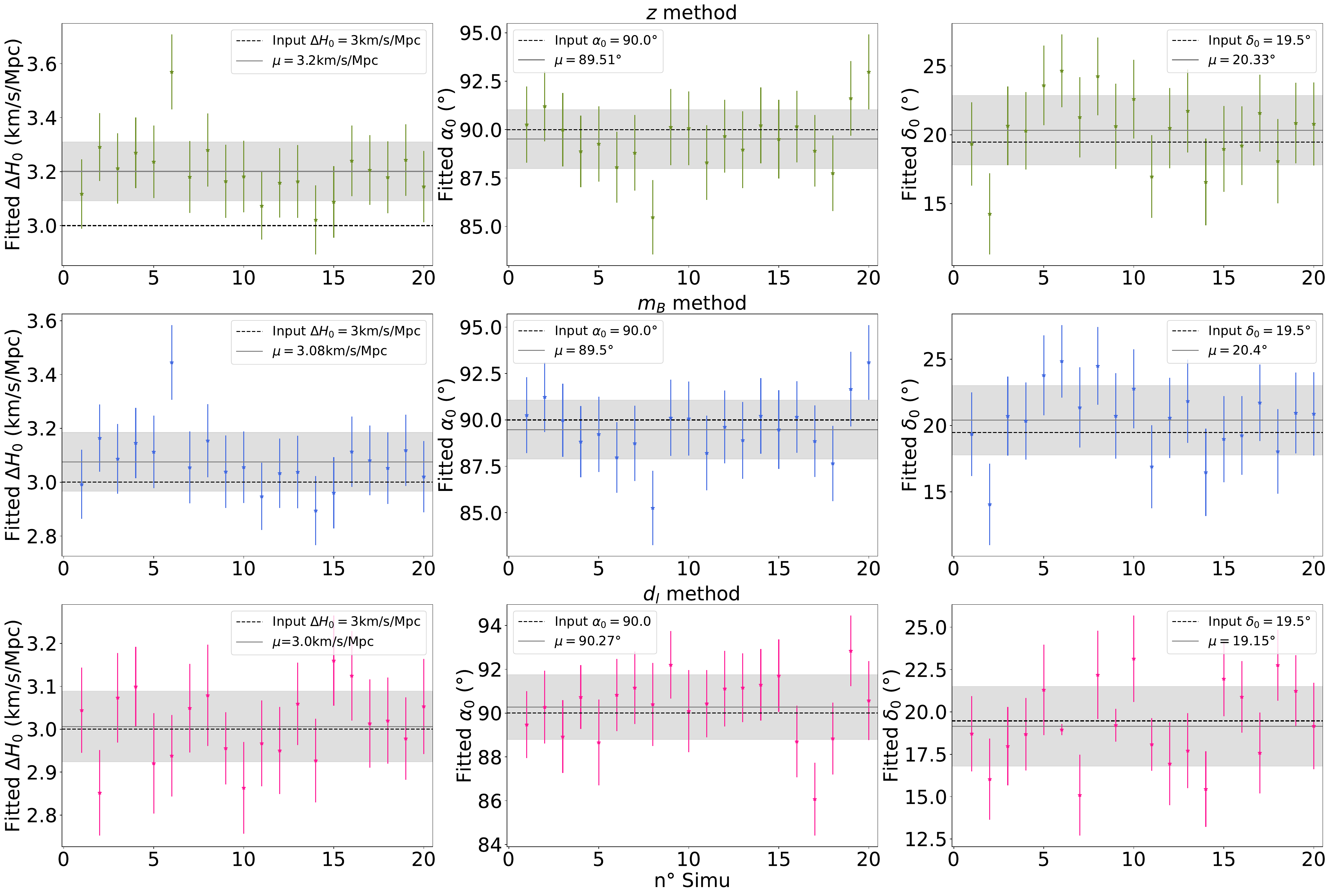}
      \caption{Fit for the $z$-method on the top panel, the $m_B$-method on the middle, and the $d_l$-method on the bottom. Left to right, the result for the 20 simulations for the dipole amplitude, $\Delta H_0$, the right ascension, $\alpha_0$, and the declination, $\delta_0$. In each panel, the initial value is represented with the black dashed line, and the median of the twenty simulation fits is in the gray line with the $1\sigma$ error represented by the gray band. For those fits, the SNe Ia colors and stretches are the ones input in the simulations.}
    \label{fig:Compa_fit_3methode}
\end{figure*}

The fit result for an input dipole of amplitude $\Delta H_0 = 3~\hunit$ and a direction $\alpha_0=90.0^\circ$ and $\delta_0=19.5^\circ$ is shown in the skymap representation for the $m_B$-method in Fig.~\ref{fig:skymap_mb}. The orange triangle correspond to the 20 fit for the 20 different simulations, the results are close to the true value in black stars. We show the fitted value for the amplitude and direction with the three dipole introduction methods in Fig.~\ref{fig:Compa_fit_3methode}. For the three methods, the true value of the location ($\alpha_0$, $\delta_0$) is contained in the $1\sigma$ uncertainties considering the median absolute deviation (MAD) of the 20 independent simulations, represented by a gray band in the plot. For the amplitude $\Delta H_0$, the $z$-method is giving a bias of $\approx 6\%$ for this example of dipole position. We interpret this bias to be caused by the low-redshift approximation of the Hubble law, which starts to deviate by $4\%$ in luminosity distance from $\Lambda$CDM at $z=0.06$.

\begin{figure}
   \centering
   \includegraphics[width=0.9\columnwidth]{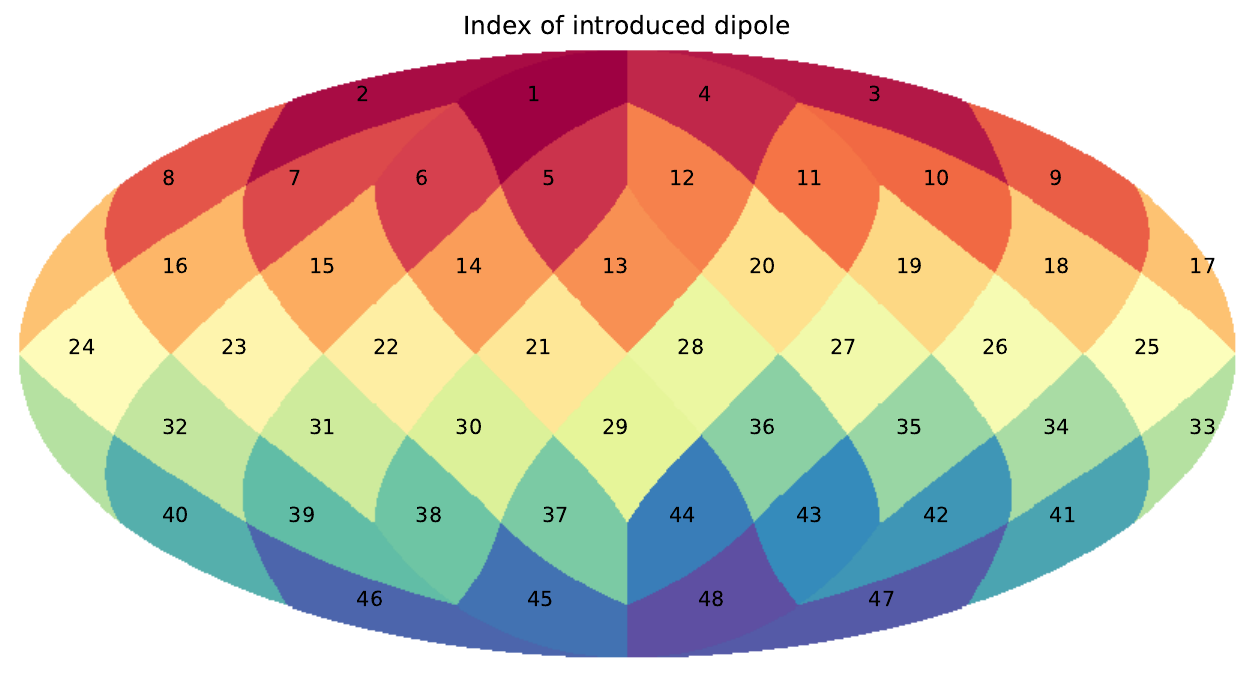}
      \caption{Skymap with the number of the initial dipole introduced. We used the center of each different patches here to obtained the 48 different dipole location. The colour of each patch allows to differentiate the 48 different locations of dipole.}
    \label{fig:indic_dip}
\end{figure}

To quantify the impact of the introduced dipole direction, we perform the same procedure for 48 different dipole positions. We evenly distribute those positions by cutting the sky with a Healpix scheme ($n_{\rm side} = 2$), as shown in Fig.~\ref{fig:indic_dip}. We used the center of the 48 patches as input dipole directions, each tested with the three introduction methods. Similarly, we use for every direction an amplitude of $\Delta H_0=3~\hunit$.

\begin{figure*}
   \centering
   \includegraphics[width=1.8\columnwidth]{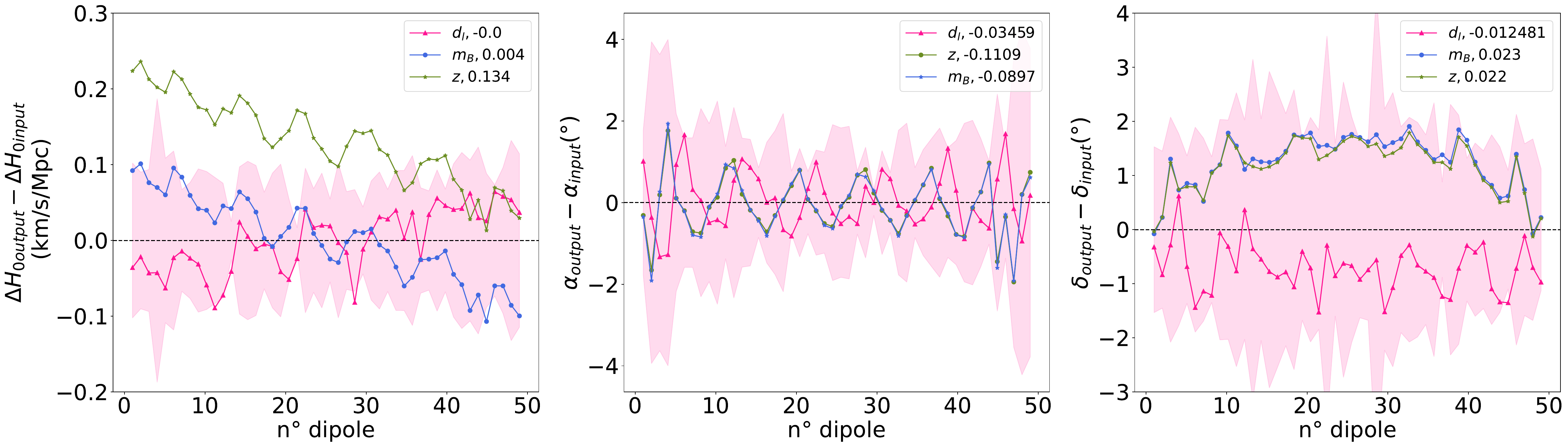}
      \caption{Difference between the output of the fit and the true value. From left to right for the dipole amplitude, the $\alpha_0$, and the $\delta_0$ directions of the dipole. The green line is for the $z$-method introduced in Sec.~\ref{sec:methodo_dip_z}, in blue for the $m_B$-method introduced in Sec.~\ref{sec:methodo_dip_mb}, and in pink for the $d_l$-method. The pink band corresponds to the typical error for the three methods. The medians are done in the label for the three different methods. As for Fig.~\ref{fig:Compa_fit_3methode}, the SNe Ia colors and stretches are the ones input in the simulations.}
    \label{fig:dip_dl}
\end{figure*}

The results for the 48 introduced dipoles are shown in Fig.~\ref{fig:dip_dl}. The green stars represent the $z$-method, the blue triangle represents the $m_B$-method, and the pink dot represents the $d_l$-method. We represent the median of the 20 simulations for each dipole. We can observe a slope in the amplitude for the three methods. Unlike the two other methods, the $d_l$-method shows a different slope direction. However, the $m_B$ and $d_l$-methods amplitudes are still contained in the error bar represented by the pink band and calculated from the median absolute deviation (MAD) over the different simulations. We use the MAD instead of the standard deviation to reduce the influence of outliers, and for consistency with the median estimator used to compute the difference between the output and input. The $z$-method produces a large bias in amplitude. Taking the median of the 48 dipole direction for the three methods, the $z$-method gives an average bias of $0.134~\hunit$ (equivalent to $4.5\%$) compared to the $m_B$-method, which is $0.004~\hunit$ (equivalent to $0.1\%$). This result confirms our previous conclusion about the non-robustness of the $z$-method compared to the other two. This method is used in the literature, such as \cite{Kalbouneh_2023}, but it is important to note that this approximation is valid only at very low redshifts and deviates from Eq.~\ref{eq:dl} with a $4\%$ effect at redshift of 0.06, for example. The intent of using this method here is to estimate the bias introduced in a quantitative manner. The results found are in agreement with the simplified calculation showing that Hubble’s law differs by up to 4\% from the luminosity distance predicted by the $\Lambda$CDM model for redshifts up to 0.06. For the $d_l$-method, we must generate as many simulations as the introduced dipole, making it computationally expensive. The $m_B$-method is a good compromise between an unbiased fit and numerical efficiency. For this reason, we use this method from now on.  

\begin{figure}
   \centering
   \includegraphics[width=0.9\columnwidth]{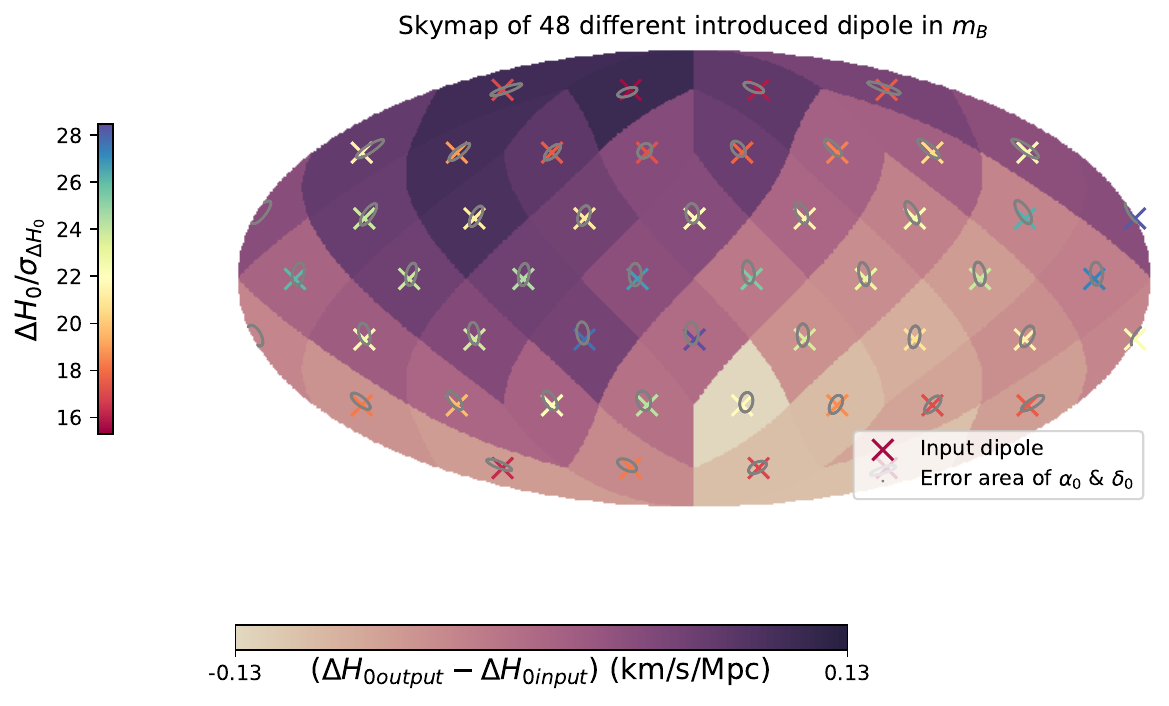}
      \caption{Skymap of the 48 different dipole fits, with the $m_B$-method, for 20 independent simulations, and using the SALT color and stretch parameters. Each cross corresponds to the input location of the 48 dipoles, and is colored according to the median of the 20 simulations sensitivity, as shown with the colorbar on the left. Each cross is associated with a patch which is colored according to the median to the difference between the input (here $3~\hunit$) and output amplitude of the dipole. All the gray ellipses are centered on the median of the different simulation fits, the ellipses represent the $1\sigma$ deviations of the fitted direction ($\sigma_{\alpha_0},\sigma_{\delta_0}$).}
    \label{fig:48_simu_mb}
\end{figure}

Fig.~\ref{fig:48_simu_mb} shows the fit results for the 20 simulations and 48 input dipole positions, but using the $x_1$ and $c$ parameters fitted by SALT instead of the generated ones. Each patch is colour-coded by the median of the difference between the output of the fit and the input of the amplitude ($3~\hunit$ for the 48 introduced dipole). We observe that the dipole amplitude is underestimated at the bottom right and overestimated at the top left by an order of $0.1~\hunit$, meaning we have more difficulty fitting the dipole near the equatorial poles. These results are visible as a slope in Fig.~\ref{fig:dip_dl}. This slope is visible for the three different methods, and its origin is investigated in more detail in Sec.~\ref{sec:sky_coverage} and seems to be caused by the ZTF cadence of observations.

In Fig.~\ref{fig:48_simu_mb}, the gray ellipses are centered around the median of the 20 fitted location dipole, and the crosses represent the position of the introduced dipole. All the crosses fall within the ellipses, which represent the $1\sigma$ uncertainty, which confirms that the true value of the location is always within the $1\sigma$ error range, whatever is the direction. The crosses are colored as a function of the median of their sensitivity ($\Delta H_0 / \sigma_{\Delta H_0}$), which is lower near the pole of the map $\Delta H_0 / \sigma_{\Delta H_0} =15$. It implies an increased difficulty in retrieving the amplitude on this side of the map, mainly caused by an increase in the amplitude uncertainty.

To estimate an error bar associated with our measurement, we choose to remain conservative by keeping the largest uncertainties associated with the MAD of the 20 simulations. It gives $\sigma_{\Delta H_0}=0.19~\hunit$ for the amplitude, $\sigma_{\alpha_0}=8.8^\circ$ for the right-ascension and $\sigma_{\delta_0}=3.9^\circ$ for the declination. With a more optimistic error bar estimation, taking the median of the MAD, we obtain respectively $0.11~\hunit$, $1.9^\circ$, and $2.9^\circ$. Even with the conservative estimator, we largely recover the amplitude of the initial dipole introduced in the simulations: $\Delta H_0 = 3.01\pm0.19~\hunit$.

\section{Systematic studies and parameter variations}
\label{sec:systematic}

In this section, we investigate different systematics which can impact our interpretations and we vary the introduced parameters.

\subsection{Varying the Hubble constant}
\label{sec:modif_of_H0}

\begin{figure*}
   \centering
   \includegraphics[width=1.8\columnwidth]{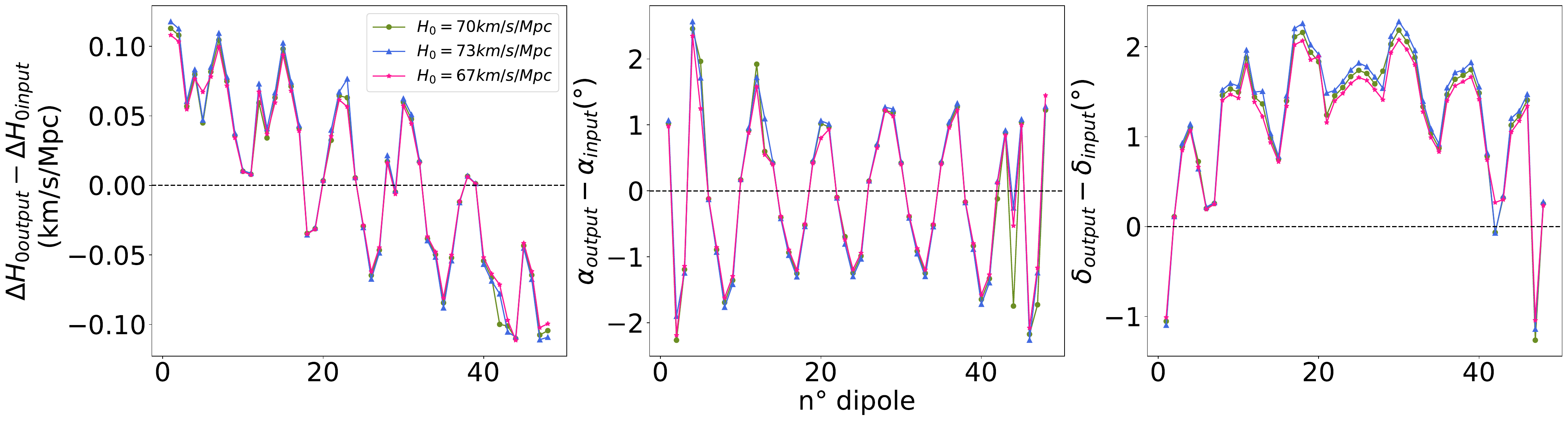}
      \caption{Comparison of the dipole fit with three different values of $H_0$ in input: $H_0=67~\hunit$ in pink stars, $H_0=70~\hunit$ in green dots, and $H_0=73~\hunit$ in blue triangles. The top panels show the comparison between the output of the fit and the input as a function of the introduced dipole; each number corresponds to the numbered position in Fig.~\ref{fig:indic_dip}.}
         \label{fig:diff_H0}
\end{figure*}

In the previous sections, we measured the dipole values from 48 different orientations, but assuming the Planck18 $H_0$ value. As mentioned in the Sec.~\ref{sec:intro}, the Hubble constant value is debated. Thus, we perform our analysis considering different $H_0$ values in the fit, while we keep the $H_0$ value of the simulation fixed. The objective is to check that our analysis is not sensitive to the choice of the $H_0$ value assumed in the fit. The results are shown in Fig.~\ref{fig:diff_H0}. We can observe that the behavior is the same between the three values. The error are exactly the same for the three different values. We conclude that our fitting methodology is not impacted by the choice of $H_0$ value in the fit.

\subsection{Sky Coverage}
\label{sec:sky_coverage}

\begin{figure*}
    \centering
    \includegraphics[width=1.7\columnwidth]{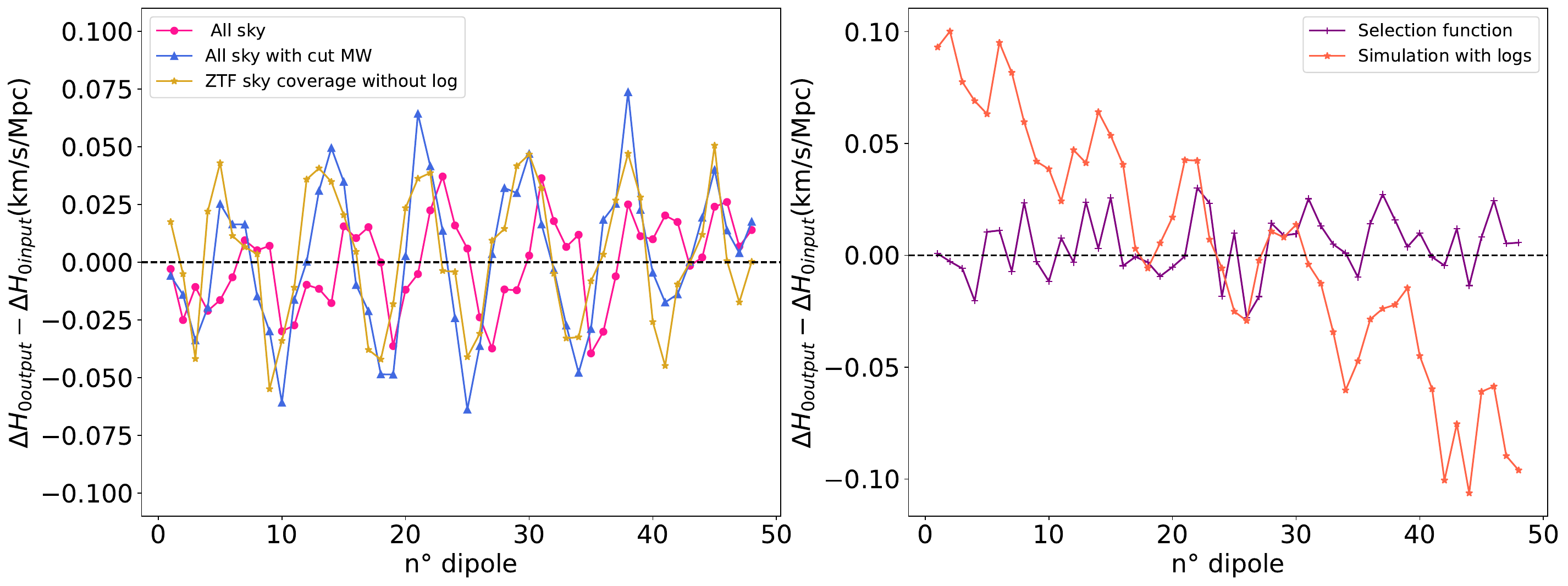}
    \caption{Difference between the output of the dipole amplitude and its input value ($3~\hunit$) as a function of the number of the dipole introduced. The different lines show the simulations with increased complexity: (left) All-sky coverage in pink, cutting Milky Way in blue, cutting in addition the south sky in yellow, (right) adding selection function effects in purple, and adding the observation logs in orange.}
    \label{fig:sky_coverage}
\end{figure*}

We analyzed different sky coverage scenarios to check for any bias introduced by the SN Ia distribution. Indeed, even if ZTF covers a large part of the sky, its anisotropic sky coverage could introduce bias. We create five new sets of simulations, keeping the same statistics as in the ZTF SN Ia DR2 ($\approx 3,600$ SNe Ia):
\begin{itemize}
    \item one with an isotropic distribution of SNe Ia in the sky without cut on the Milky Way (MW). Represented in deep pink dot in the Fig.~\ref{fig:sky_coverage}. 
    \item one with a full sky but a cut on the MW represented in blue in Fig.~\ref{fig:sky_coverage}
    \item another with a cut in MW and below $-30^\circ$ in declination to reproduce the ZTF sky coverage but without any SN Ia selection cut, represented in yellow in Fig.~\ref{fig:sky_coverage}.
    \item one with a cut in MW and a cut below $-30^\circ$ in declination to reproduce the ZTF sky coverage, and with the function selection that reproduces the spectroscopic follow-up of BTS, as explained in Sec.~\ref{sec:ZTF_DR2}, in purple in Fig.~\ref{fig:sky_coverage}
    \item a last one with the ZTF skymap and the observing logs containing all the observation information, in orange in Fig.~\ref{fig:sky_coverage}.
\end{itemize}

As we focus only on testing the impact of the footprint, we choose to use the stretch $x_1$, colour $c$, and $x_0$ generated parameters in the fit of those new simulations. For each set of sky coverage, we simulate 20 independent simulations, introduce the dipole with the $m_B$-method, and fit it with the same procedure as in previous sections. Fig.~\ref{fig:sky_coverage} shows the difference between output and input for the amplitude of the dipole; the right ascension and declination are shown in Appendix~\ref{appx:sky_coverage}.

If we consider the full sky, the MW cut, and the cut below $-30^{\circ}$, the amplitude is unbiased, as in \cite{Kalbouneh_2025}. The coverage of the sky does not seem to introduce a bias. However, the error for the anisotropic case is slightly bigger: $\sigma_{\Delta_{H_0}}=0.073~\hunit$, $\sigma_{\Delta_{H_0}}=0.090~\hunit$, respectively for the whole sky and anisotropic sky.

The slope in Sec.~\ref{sec:Dipole_analyze} in the amplitude appears only when we add the effect of the observing log, which contains possible technical issues and the cadence of observations. We interpret that the difficulty in recovering the signal for the amplitude at the pole comes from the cadence of the observation and not from the lack of observations below $-30^{\circ}$. We also observe oscillations for all different sky coverage scenarios as a function of the initial location of the dipole. These residual oscillations seem to be caused by unexplained small correlations between $M_B$ and $\alpha_0$ parameters for the whole sky, and amplified by a correlation between $\alpha_0$ and $\delta_0$ for the realistic ZTF simulations. These oscillations will be investigated further in a future analysis with the data. However, we bring more elements in their potential origin in Appendix ~\ref{appx:sky_coverage}. In spite of their presence, since they remain well below the median error of the simulations, we consider this residual oscillation as negligible.

\subsection{Large-scale structure impact}
\label{sec:large_scale_strucutre}

\begin{figure*}
   \centering
   \includegraphics[width=1.8\columnwidth]{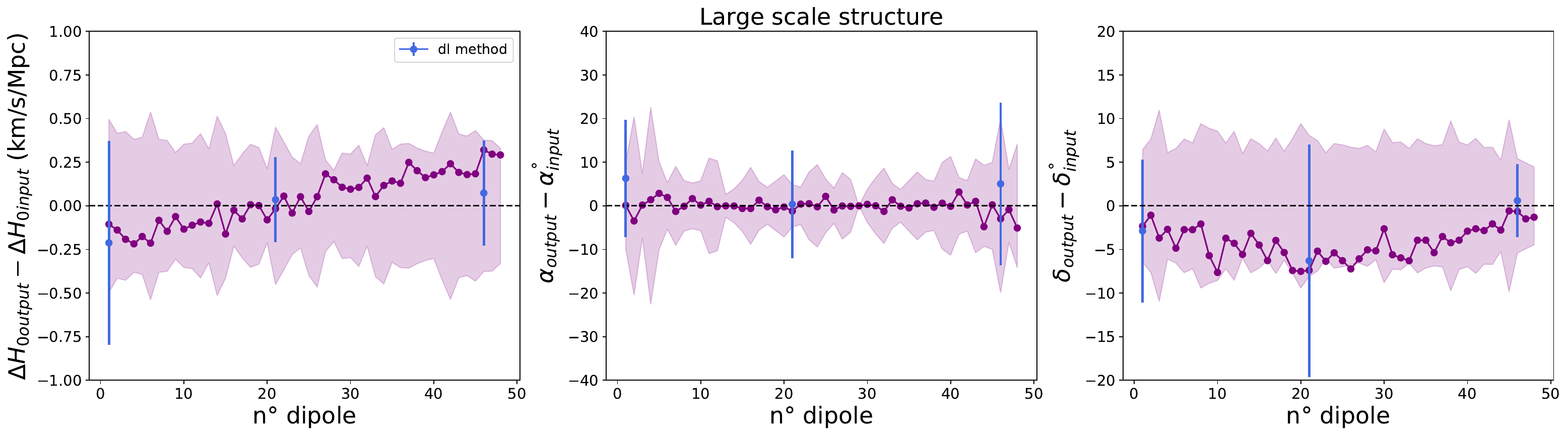}
      \caption{Difference between the output of the fit and the truth value, for 27 different simulations with the large-scale structures introduced. Left to right, the figure shows the dipole amplitude, the RA, and the DEC. The purple band is the MAD of the 27 simulations centred on zero. The blue dots correspond to the median over 27 simulations with the initial dipole introduced with the $d_l$-method, with the associated error computed which is the MAD of the 27 simulations, for an initial amplitude of $3~\hunit$.}
    \label{fig:large_scale_structure}
\end{figure*}

The peculiar velocities shift the observed redshift of SNe Ia \citep{Carreres_2023}. Including the effect of peculiar velocities, generated from a realistic N-body simulation, allows to reproduce the ZTF SN Ia DR2.5 dataset more realistically. With the large-scale structure implemented, we run 27 independent simulations. We reproduce the same method as mentioned in Sec.~\ref{sec:Dipole_analyze} for the $m_B$-method and for the 48 different introduced dipoles, the results are shown in Fig.~\ref{fig:large_scale_structure}. We applied a cut on redshift $z>0.02$, nearby SNe Ia for which the impact of PV is too important and could introduce a bias in the results, as we would do for the real dataset.

For the three dipole parameters, the amplitude and location, we recover the initial value for the 48 different dipoles. As seen in Fig.~\ref{fig:large_scale_structure}, the difference between the fitted value and the introduced one is included in the $1\sigma$ uncertainties. Even if we see a slight offset of $\approx 4.1^\circ$ for the declination, the curve remains included in the uncertainty band. This demonstrates the robustness of the method in recovering the parameters.

For an initial amplitude of $3~\hunit$, we estimate an amplitude of $3.05\pm0.33~\hunit$. We still have a trend for the difference in the amplitude, but in the opposite direction as compared to the simulations without large-scale structure in Sec.~\ref{sec:Dipole_analyze}. The Fig.~\ref{fig:large_scale_structure} displays the median values of the 27 different simulations, but considering the simulations one by one, the direction of this slope can be different. Although the orientation of the slope varies from one simulation to another, the slope consistently appears. This indicates that it is unlikely to be a statistical effect and more likely driven by the impact of the ZTF cadence.

We tested introducing the two extreme cases of the slope as well as a point at the center of the slope using the $d_l$-method. These dipoles correspond to numbers 1, 21 and 46 in Fig.~\ref{fig:indic_dip}. The objective here was to determine whether the slope would also have been observed using the $d_l$ method, which is the most robust but more computationally intensive. The medians of the 27 simulations for the three cases are shown in blue with their associated errors in Fig.~\ref{fig:large_scale_structure}. We note that the positions of the medians correspond to those obtained with the $m_B$-method and that the errors are similar, which is expected since this depends mainly on the statistics of the analysis. Thanks to these three points at the ends and in the center, we can see that the slope is also visible using the $d_l$ method. This result is not surprising, given the conclusions presented in Sec.~\ref{sec:sky_coverage}, where we observed that this slope appeared to originate from the observational logs. We also compared the simulation results obtained using the two methods for the three parameters. The median per-realisation $m_B$ - $d_l$ amplitude offsets are 0.14, 0.09 and 0.02 \hunit, with uncertainties of about 0.11, 0.09 and 0.12 \hunit (define as nMAD/$\sqrt{N_{\text{Simu}}}$), where each realisation corresponds to a pair of simulations sharing the same large-scale structure box. These results confirm that the $m_B$ and $d_l$-methods are similar and that the systematic analysis performed with the $m_B$-method does not appear to introduce any bias. 

\subsection{Volume-limited}
\label{sec:v_lim}

\begin{figure*}
   \centering
   \includegraphics[width=1.8\columnwidth]{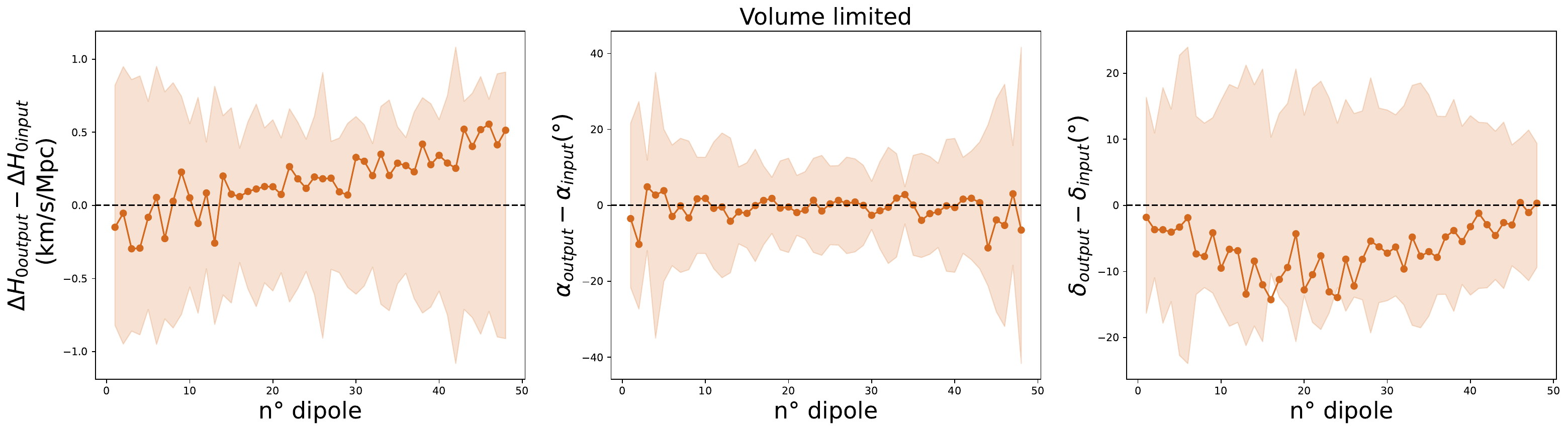}
      \caption{Same as Fig.~\ref{fig:large_scale_structure} for the volume-limited sample.}
 \label{fig:V_lim}
\end{figure*}

We also tested our method on a volume-limited sample, composed of $\approx 1000$ SNe Ia with a redshift cut $z<0.06$ following~\citep{Amenouche_2025}. It is an unbiased sample where the intrinsic distribution of SN Ia is complete, which means we are less sensitive to the Malmquist bias. Fig.~\ref{fig:V_lim} shows the results of the same fit on the large-scale structure simulations, but with the volume-limited sample. Similarly, we also apply the redshift cut $z>0.02$.

The errors are bigger due to a smaller statistic, as we can see in Tab.~\ref{tab:sens_error}. In addition, we have a bigger offset for the difference between the input declination and output, of $6.5^\circ$, but still consistent with the $1\sigma$ uncertainties. We conclude that using the volume-limited sample allows to recover the introduced dipole, but with an increase in the declination offset and in the statistical uncertainty.

Moreover, the small bias observed in the dipole amplitude and declination as a function of the dipole direction has the same trend when comparing the dipole fit result between the volume-limited sample (Fig.~\ref{fig:V_lim}) and the full redshift sample (Fig.~\ref{fig:large_scale_structure}), indicating that the observed bias is not related to any selection (Malmquist bias) effect, but to the observing cadence as already discussed in section~\ref{sec:sky_coverage}. And beyond the volume-limited sample $z<0.06$, it was shown in~\citep{Amenouche_2025} that the sub-sample $z<0.08$ (representing almost 2/3 of the total ZTF SN Ia DR2 sample) is marginally biased when standard cosmological cuts are applied on SN Ia color parameter. So, our interpretation is that the dipole fit is driven mostly by a sub-sample almost not affected by selection effects.

The volume-limited sample allows addressing a concern about the introduction of a dipole via the $m_B$-method. Indeed, we can naively expect that the introduction of a dipole via the physical $d_l$-method should impact the magnitude and then affect the full sample in terms of redshift distribution, while with the $m_B$-method the sample is frozen before introducing the dipole, whatever is its direction. But in fact, the effect would be more to change marginally the SNe Ia repartition in the sky than the full redshift distribution because we can expect to lose SNe Ia in the direction of the dipole, and to gain SN in the opposite direction. So, this will affect more the SN repartition than the redshift distribution. But as the volume-limited sample is not magnitude limited, it is not affected by this effect. And as the results obtained with the full redshift sample are similar to those obtained with the volume-limited sample, we can conclude that the $m_B$-method is not biasing significantly the sample.

\subsection{Varying the amplitude}
\label{sec:modif_of_DH0}

We aim to investigate the dependencies of our study with respect to the input dipole amplitude. We perform the same analysis as before with the following values: $\Delta H_0=2~\hunit$ and $\Delta H_0=1~\hunit$, for the 27 simulations containing the large-scale structure. 

\begin{figure*}
   \centering
   \includegraphics[width=1.8\columnwidth]{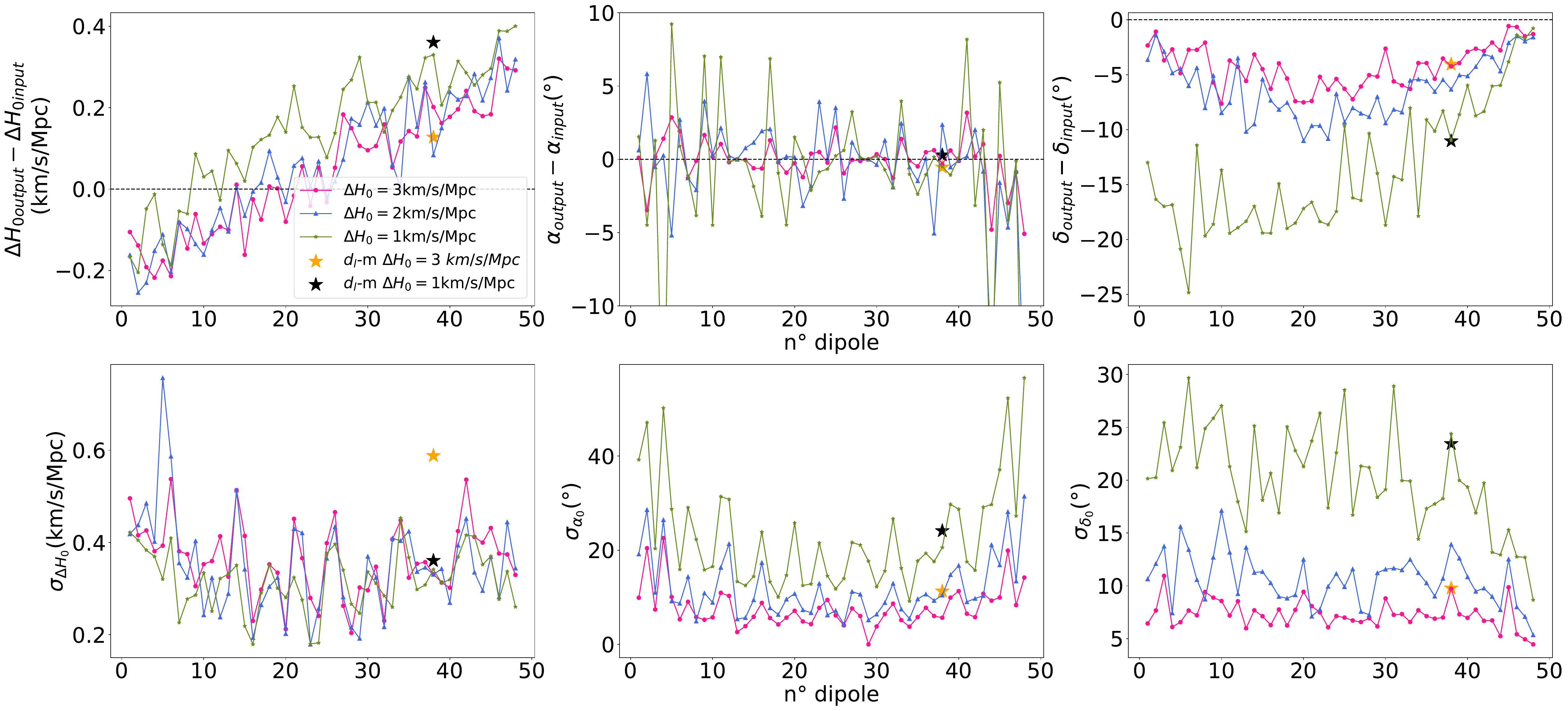}
      \caption{Comparison of three different values of amplitude of the dipole $\Delta H_0$ in input: $\Delta H_0=3~\hunit$ in pink dot, $\Delta H_0=2~\hunit$ in blue triangles and $\Delta H_0=1~\hunit$ in connected green stars. The representation of the results is the same as in Fig.~\ref{fig:diff_H0}. The black and yellow stars represent tests conducted using the $d_l$ method for a different amplitude in order to determine whether a difference was observed between the two methods.}
\label{fig:diff_DH0}
\end{figure*}

Fig.~\ref{fig:diff_DH0} shows the difference between the input and output of the three parameters, and the error of those parameters computed from the MAD. As expected, it is harder to recover the location of the initial dipole introduced for an amplitude of $1~\hunit$. Indeed, the offset in $\delta_0$ is more pronounced for the smaller amplitude, and the amplitude of the oscillations in RA is larger. Moreover, the error for this case is more significant than for the other two amplitudes for the dipole position. Focusing on the dipole amplitude in Fig.~\ref{fig:diff_DH0}, the error and the difference between the input and output are the same for the three different values. This result confirms the robustness of our analysis and that we can recover a dipole even with an amplitude of $1~\hunit$, which approximately corresponds to the $H_0$ error in the direct measurement \citep{Riess_2022}.

\subsection{Anisotropies study without dipole incorporation in the simulations}
\label{sec:without_dip}

\begin{figure}
   \centering
   \includegraphics[width=0.9\columnwidth]{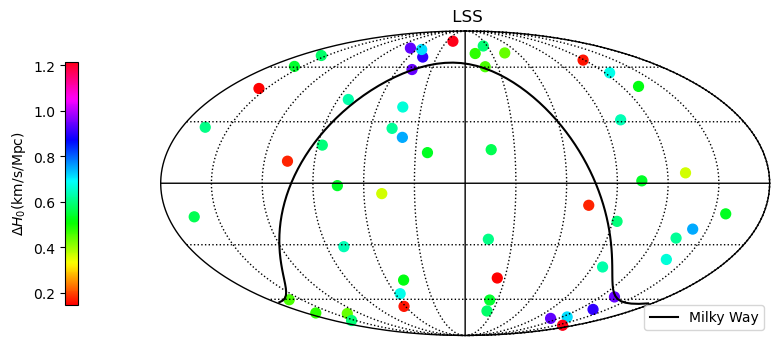}
   \includegraphics[width=0.9\columnwidth]{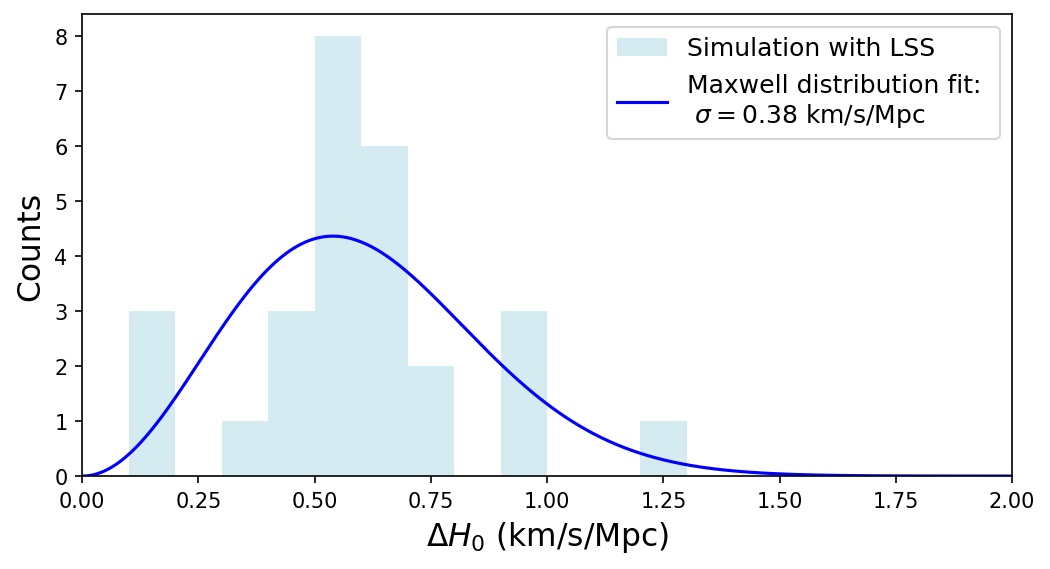}
      \caption{Upper plot: Skymap of the 27 different fit on simulations without any dipole introduced. Each dots is coloured depending the values of the dipole amplitude. Lower plot: $\Delta H_0$ distribution of the 27 simulations (light blue histogram) and the fitted result of a Maxwell distribution (blue) giving $\sigma = 0.38~\hunit$.}
\label{fig:large_scale_structure_without_dipole}
\end{figure}

We test our methodology in the case when no dipole is introduced in the 27 large-scale structure simulations. We obtain 27 different dipole locations and amplitudes reported in Fig.~\ref{fig:large_scale_structure_without_dipole}. The highest amplitudes of dipole obtained is $1.2~\hunit$. However, the MAD for this case is only $0.12~\hunit$, lower than the one obtained for the three different dipole amplitudes introduced in the simulations in Sec.~\ref{sec:modif_of_DH0} ($\approx 0.34~\hunit$). The difference between those values results from a systematic error in the initial position of the introduced dipole. Indeed, the slope present for the three different introduced amplitudes in Fig.~\ref{fig:diff_DH0} induces a systematic bias as a function of the initial position of the introduced dipole. We have to take it into account in the total error budget. 

Our error model combines a statistical term, the median error of the 27 fits with no dipole in the simulations $\sigma_{\rm stat}=0.16~\hunit$, with a systematic term derived from the amplitude slope. To quantify the systematic error, we fit a linear function of the maximum of the slope for the three introduced amplitudes ($1~\hunit$,$2~\hunit$,$3~\hunit$): $\sigma_{\rm sys}=f(\Delta H_0)$. We estimate the total error from the quadratic sum: $\sigma_{\rm tot}=\sqrt{\sigma_{\rm stat}^2+\sigma_{\rm sys}^2}$. We use the same schema to build an error model for $\alpha_0$ and $\delta_0$.

We use this error model to compute the error of the previous different analyses in Sec.~\ref{sec:systematic}. These errors are summarized in Tab.~\ref{tab:sens_error} with their sensitivity. Concerning the result without introducing a dipole in the simulation, for the systematic error we use our linear function extrapolated to a zero amplitude: $\sigma_{\rm sys}=0.44~\hunit$. At the end we have a total error of: $\sigma_{\rm tot}=0.47~\hunit$ for the amplitude, $24^\circ$ for $\alpha_0$ and $34^\circ$ for $\delta_0$. The median value of the fitted amplitudes for the 27 different simulations is $0.59~\hunit$ for the full redshift volume of ZTF. 
From a statistical point-of-view, this null test dipole corresponds to a zero-mean isotropic Gaussian dipole with a dispersion $\sigma$ per component. In this case the expected amplitude $r \equiv \Delta H_0$ (positive-definite) is Maxwell-distributed  
\begin{equation}
\label{eq:Maxwell}
      f(r)=\sqrt{\frac{2}{\pi}}\frac{r^2}{\sigma^3}\exp\left(-\frac{r^2}{2\sigma^2}\right),
\end{equation}
characterized by a median equal to $1.54 \sigma$. The amplitude distribution (light-blue histogram in Fig.~\ref{fig:large_scale_structure_without_dipole}, lower plot) corresponds to a Maxwell-distribution with the fitted dispersion $\sigma = 0.38$, for which the median is in agreement with the observed one $0.59~\hunit$. This means that the statistical uncertainty $\sigma_{\rm stat}=0.16~\hunit$ is not representative of the Maxwell-distribution dispersion. This observation points to an underestimation of the uncertainty with missing components like the sample variance of the peculiar velocities.

\begin{figure}
   \centering
   \includegraphics[width=0.9\columnwidth]{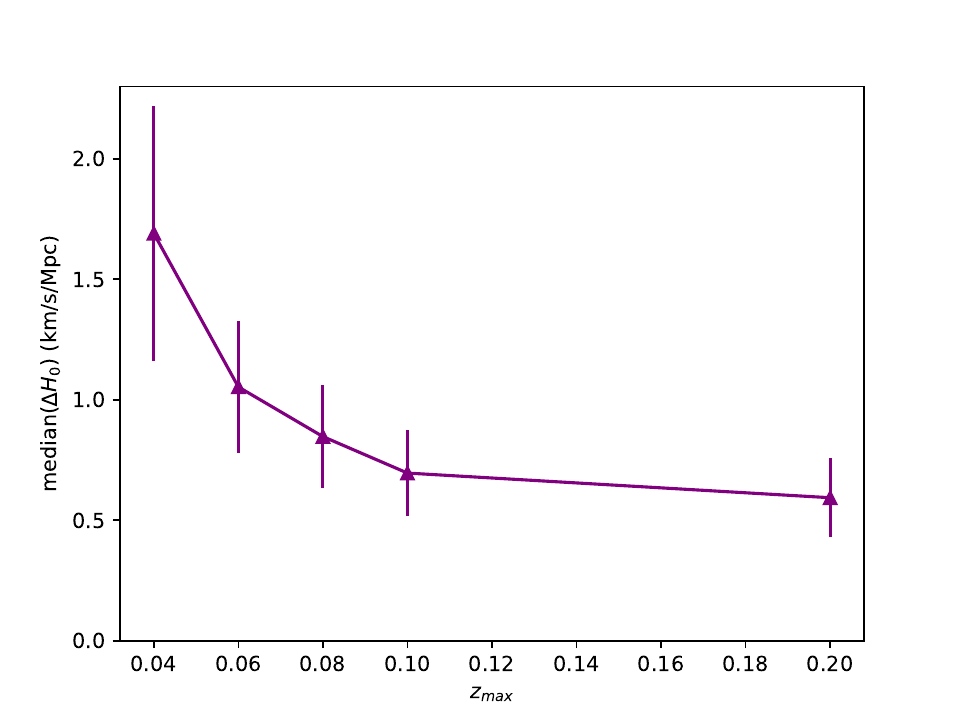}
      \caption{Median of the 27 simulations of the dipole amplitude without introducing the dipole with the median of the fitting errors, as a function of different subsamples cut at different redshift.}
    \label{fig:Diff_cut}
\end{figure}

In these simulations, no bulk motion has been included. This means that the observed amplitude is due to the large-scale structure imprinting a residual pattern that mimics a dipole. If this is true, then the fitted amplitude should be higher over a smaller volume. A study was carried out on these same simulations by applying different redshift cuts. Indeed, with a cut-off where we retain only those SNe Ia with a redshift below $0.04$, we observe an average amplitude across the 27 simulations of $1.69~\hunit$ as shown in Fig.~\ref{fig:Diff_cut}. The average fitting amplitude decreases as the redshift volume increases. This decrease leads us to speculate that the remaining amplitude of $0.59~\hunit$ for the full sample could be due to a large-scale structure imprint in the simulations which correspond to the sample variance of the peculiar velocities in each sub-box, plus noise. A larger volume allows to no longer observe this possible large-scale structure imprint and to isolate the observation of a dipole on $H_0$. The inclusion of the large-scale structure in the likelihood covariance matrix to assess its impact on our uncertainties is also left for future study when the methodology will be applied to real data.

In any case, for future real-data analysis, the measured dipole amplitude will have to be confronted to the null test hypothesis (no dipole) following a Maxwell-distribution. The 95th percentile of the null-fit amplitude distribution is $1.06~\hunit$, meaning that a dipole amplitude measured below this value on real data would be consistent with the null hypothesis.

\begin{table}[h!]
\centering
\begin{tabular}{lcccc}
\toprule
\textbf{Simulation} & $\sigma_{\Delta H_0}$ & $\frac{\Delta H_0}{\sigma_{H_0}}$ & $\sigma_{\alpha_0}$ & $\sigma_{\delta_0}$ \\
\midrule

\multirow{1}{*}{$\Delta H_{00}$}  & $\pm0.47$ & $1$ & $\pm24$ & $\pm34$\\

\midrule
\multirow{1}{*}{$\Delta H_{01}$}  & $\pm0.41$ & $3$ & $\pm10$ & $\pm23$\\

\midrule
\multirow{1}{*}{$\Delta H_{02}$} & $\pm0.37$ & $6$ & $\pm6.4$ & $\pm14$\\

\midrule
\multirow{1}{*}{$\Delta H_{03}$}  & $\pm0.33$ & $9$ & $\pm3.4$ & $\pm6.1$\\

\midrule

\multirow{1}{*}{$\Delta H_{03} (V_{\rm lim})$}  & $\pm0.42$ & $ 3 $ & $\pm17$ & $\pm15$\\

\bottomrule
\end{tabular}

\caption{Sensitivity and error for the different cases analyzed, for the three free parameters of the dipole: the amplitude $\Delta H_0$ in $\hunit$, the location with $\alpha_0$ and $\delta_0$ in degrees. Here, the sensitivity is the value over the error of the value, with the error built with the error model described in Sec.~\ref{sec:without_dip}. $\Delta H_{01}$,$\Delta H_{02}$, $\Delta H_{03}$ correspond to an initial amplitude introduced equal to $1~\hunit$, $2~\hunit$, $3~\hunit$ mentioned in Sec.~\ref{sec:modif_of_DH0} with the simulations containing the large-scale structure. The $\Delta H_{00}$ corresponds to the analyses without initial dipole introduced in the simulation discussed in Sec.~\ref{sec:without_dip}. $V_{\rm lim}$ corresponds to the analysis with the volume-limited sample detailed in Sec.\ref{sec:v_lim} for an initial amplitude of $3~\hunit$.}
\label{tab:sens_error}
\end{table}


\section{Discussion}
\label{sec:discussion}

This analysis aims to build a method to measure a possible $H_0$ anisotropy and verify the accuracy of the cosmological principle. Our method focuses on a dipole-type anisotropy on $H_0$ in the ZTF SN Ia DR2.5 sample. We test for various potential systematics and analysis variation, focusing more particularly on the impact of peculiar velocities on our measurement. Indeed, the peculiar velocities caused by the gravitational pull of nearby structures can mimic a bulk flow, i.e., a non-null average velocity in the vicinity of the observer compared to the reference frame. Even with this kind of effect, we can recover a dipole with an amplitude of $3~\hunit$ on ZTF simulation with an accuracy of $0.33~\hunit$ for an initial amplitude of $3~\hunit$, and of $3.4^\circ$ and $6.1^\circ$ for the right ascension and declination as shown in Tab.~\ref{tab:sens_error}.

In Sec.~\ref{sec:without_dip}, we perform our analysis without introducing a dipole in the simulations containing large-scale structure effects. This test yields a dipole whose fitted amplitude is inconsistent with zero. This amplitude could be due to a large-scale structure imprint present in the simulations. Even if an $H_0$ anisotropy is an intrinsic property of the Universe, a $H_0$ dipole can also be interpreted as a bulk-flow effect in a defined volume. However, in these simulations, a global motion of the observer that would cause a bulk flow is not introduced. This means the residual dipole is the consequence of peculiar velocities. For future investigation with data, to be able to distinguish a bulk-flow effect from an $H_0$ dipole, an analysis at different redshifts that covers a large part of the sky will be required.

Some other studies found a $H_0$ dipole. For example, \cite{Migkas_2021} detects an $H_0$ variation of $9\%$ using galaxy clustering. This variation is equivalent to a dipole amplitude of $6~\hunit$. As we have a sensitivity of $9$ for an amplitude of $3~\hunit$ in our configuration, we could easily measure these $H_0$ variations. Other studies detected a quadrupole-type anisotropy, e.g., \cite{Cowell_2023}. In our article, we limit ourselves to the study of an $H_0$ dipole, but future studies should be extended to account for higher multipole orders.

As explained in Sec.~\ref{sec:Dipole_analyze}, we estimate the standardization parameters at the same time as the dipole parameters. A confirmation of the robustness of the analyses is our ability to recover the standardization parameters used to produce the simulations. The input of the simulations, for a full survey, was: $M_B=-19.3,~\alpha =0.14,~\beta=3.15,~\sigma_{\text{int}}=0.14$. In average for the 48 different dipole locations and 20 simulations, we obtain: $M_B=-19.313 \pm 0.004,~\alpha=0.139\pm0.003,~\beta=3.11\pm0.02,~\sigma_{\text{ int}}=0.128\pm0.001$. Regarding the errors, we do not recover the initial parameters. For $M_B$, which is degenerated with $H_0$, its value changes as a function of the $H_0$ value used in the fit. Concerning $\alpha$, $\beta$ and $\sigma_{\text{int}}$, in this analysis we do not use the complete covariance matrix, we only have the diagonal term, so their uncertainties are underestimated. 

All the figures containing the difference between the output parameter and input (see e.g., Fig.~\ref{fig:large_scale_structure}) show oscillations as a function of the sky position of the introduced dipole. We investigated potential fit issues, which could be caused by boundary effects when the fitted RA value reaches the limit. Changing the reference frame for the fit by centring it on the $\alpha_0$ value found during the first fit does not affect our results. Furthermore, modifying the fitter limit for RA for the second fit, i.e., from $[0;2\pi]$ to $[\alpha_1-\pi;\alpha_1+\pi]$ where $\alpha_1$ is the RA guessed during the first fit, has also no impact. These oscillations are included within the error, so are not likely to affect our analysis. Future work will investigate further this effect.


\section{Conclusion}
\label{sec:conclusion}

This paper details a methodology to measure anisotropy in $H_0$ in realistic ZTF SN Ia DR2.5 simulations. Our analysis is focused on a dipole anisotropy. We test three methods to introduce the dipole in our simulations; one consisting of introducing it in the redshift of the SN Ia ($z$-method), the second one introduced the dipole in the compute magnitude of the SN Ia ($m_B$-method). The last one introduces it in the luminosity distance ($d_l$-method). Across all the different positions tested, the $z$-method introduces an average bias of $\approx4.5\%$ in the recovery of the dipole amplitude. The analysis is not biased against the two other methods, but the $d_l$-method is more expensive numerically. For this reason, we used the $m_B$-method for all the analyses to implement the dipole in the simulations. 

Our conclusions concerning the analysis are the following: 

\begin{enumerate}
      \item Our measurement is affected by the observing logs, which cause a bias in declination in particular. However, the sky coverage, considering the full sky or the ZTF footprint, has no impact on our study.
      \item The fitted dipole is not dependent on the $H_0$ value, both for amplitude and direction.
      \item We build an error model containing both statistical and systematic errors, depending on the initial position of the introduced dipole.
      \item We are able to recover an anisotropy of an amplitude of $3~\hunit$, with a precision of $0.33~\hunit$ for realistic simulations containing large-scale structure. Regardless of the position of the introduced dipole, we recovered it with errors of $3.4^\circ$, $6.1^\circ$ for the right ascensions and declination for a conservative case. The fitted declination is biased $\approx 4^\circ$, but this bias is still contained in the $1\sigma$ uncertainties.
      \item For smaller dipole amplitudes such as $1~\hunit$, our errors are larger, and the sensitivity, $\Delta H_0 / \sigma_{\Delta H_0}$, decreases. However, we still recover the signal with a significance of $3\sigma$.
      \item Using the unbiased volume-limited sample reduces the sensitivity but gives similar results.
      \item We perform our analysis without introducing a dipole in the simulations and then fit a dipole. The fitted dipole amplitude is not consistent with zero, which leads us to believe that we fit a residual dipole from peculiar velocities. This residual highlights the importance of investigating the cause of the anisotropy if it is ultimately observed in the data. This study provides us with uncertainty figures that will be applied to the data: $0.47~\hunit$ for the amplitude and $24^\circ$ for the right ascension and $34^\circ$ for the declination. Any dipole amplitude measured on real data must be compared against this null distribution; its 95th percentile is $1.06~\hunit$, meaning that a measured amplitude below this value would not constitute evidence for a true $H_0$ anisotropy.
\end{enumerate}
   
This first methodology focused on a dipole analysis; another study with a higher multipole, such as a quadrupole, will be investigated in the future. Moreover, this study employs a methodology built on simulations; the application of this method to the ZTF SN Ia DR2.5 data will be the subject of a second study. 

The next data release of ZTF, named DR3, will drastically increase the number of SNe Ia to around 10,000. A planned data release with photometrically-typed SNe Ia, named DR4, will increase this number even more to 40,000. Applying our methodology to those data releases will extensively improve the sensitivity and the errors of the analysis. It is also possible to use our method on other present and future surveys such as LSST \citep{LSST_2018} or ATLAS \citep{ATLAS_2018}.

\begin{acknowledgements}
Based on observations obtained with the Samuel Oschin Telescope 48-inch and the 60-inch Telescope at the Palomar Observatory as part of the Zwicky Transient Facility project. ZTF is supported by the National Science Foundation under Grants No. AST-1440341 and AST-2034437 and a collaboration including current partners Caltech, IPAC, the Weizmann Institute of Science, the Oskar Klein Center at Stockholm University, the University of Maryland, Deutsches Elektronen-Synchrotron and Humboldt University, the TANGO Consortium of Taiwan, the University of Wisconsin at Milwaukee, Trinity College Dublin, Lawrence Livermore National Laboratories, IN2P3, University of Warwick, Ruhr University Bochum, Northwestern University and former partners the University of Washington, Los Alamos National Laboratories, and Lawrence Berkeley National Laboratories. Operations are conducted by COO, IPAC, and UW. Y.-L.K. was supported by the Lee Wonchul Fellowship, funded through the BK21 Fostering Outstanding Universities for Research (FOUR) Program (grant No. 4120200513819) and the National Research Foundation of Korea to the Center for Galaxy Evolution Research (RS-2022-NR070872, RS-2022-NR070525). U.B and T.E.M.B is funded by Horizon Europe ERC grant no. 101125877. G.D. acknowledges support from the European Union’s Horizon Europe research and innovation programme under the Marie Skłodowska-Curie grant agreement No 101199369. L.G. acknowledges financial support from AGAUR, CSIC, MCIN and AEI 10.13039/501100011033 under projects PID2023-151307NB-I00, PIE 20215AT016, and CEX2020-001058-M. MG is supported by the European Union’s Horizon 2020 research and innovation programme under ERC Grant Agreement No. 101002652 (BayeSN; PI K. Mandel). N.R. is supported by a Northwestern University Presidential Fellowship Award. Zwicky Transient Facility access for N.R. was supported by Northwestern University and the Center for Interdisciplinary Exploration and Research in Astrophysics (CIERA). A.G. acknowledges financial support from the research project grant “Understanding the Dynamic Universe” funded by the Knut and Alice Wallenberg under Dnr KAW 2018.0067, {\em Vetenskapsr\aa det}, the Swedish Research Council through grants project Dnr 2020-03444.

\end{acknowledgements}

\bibliographystyle{aa}
\bibliography{Biblio}

\begin{appendix}

\section{Correlation between fitted parameters}
\label{appex:correlation}

\begin{figure}
   \centering
   \includegraphics[width=0.9\columnwidth]{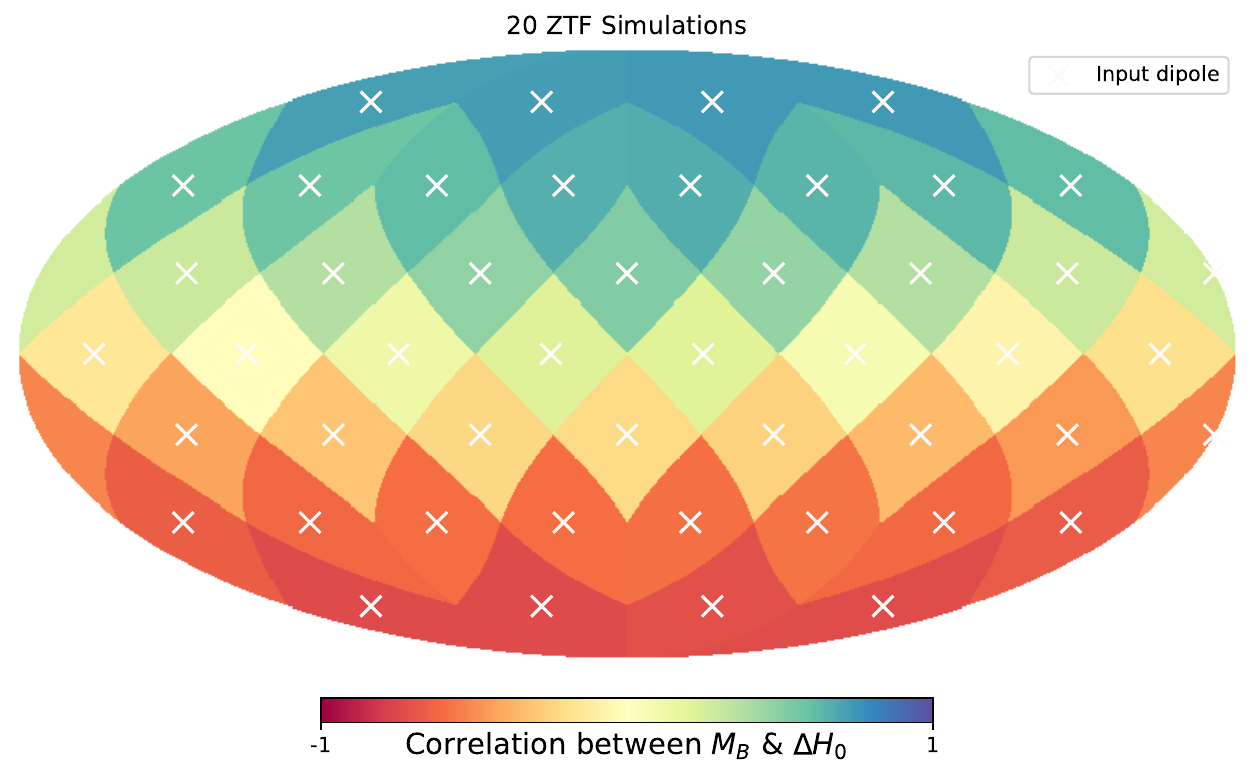}
      \caption{Skymap of the correlation between the parameter $M_B$, and the dipole amplitude, each white cross corresponds to a different location of dipole introduced in the data with the $m_B$ method, for the 20 simulations with an initial amplitude of dipole of $3~\hunit$. Each patch is colored depending of the median of the correlation between the two parameters for the 20 simulations.}
         \label{fig:annex_corr_Mb_H0}
\end{figure}

\begin{figure}
   \centering
   \includegraphics[width=0.9\columnwidth]{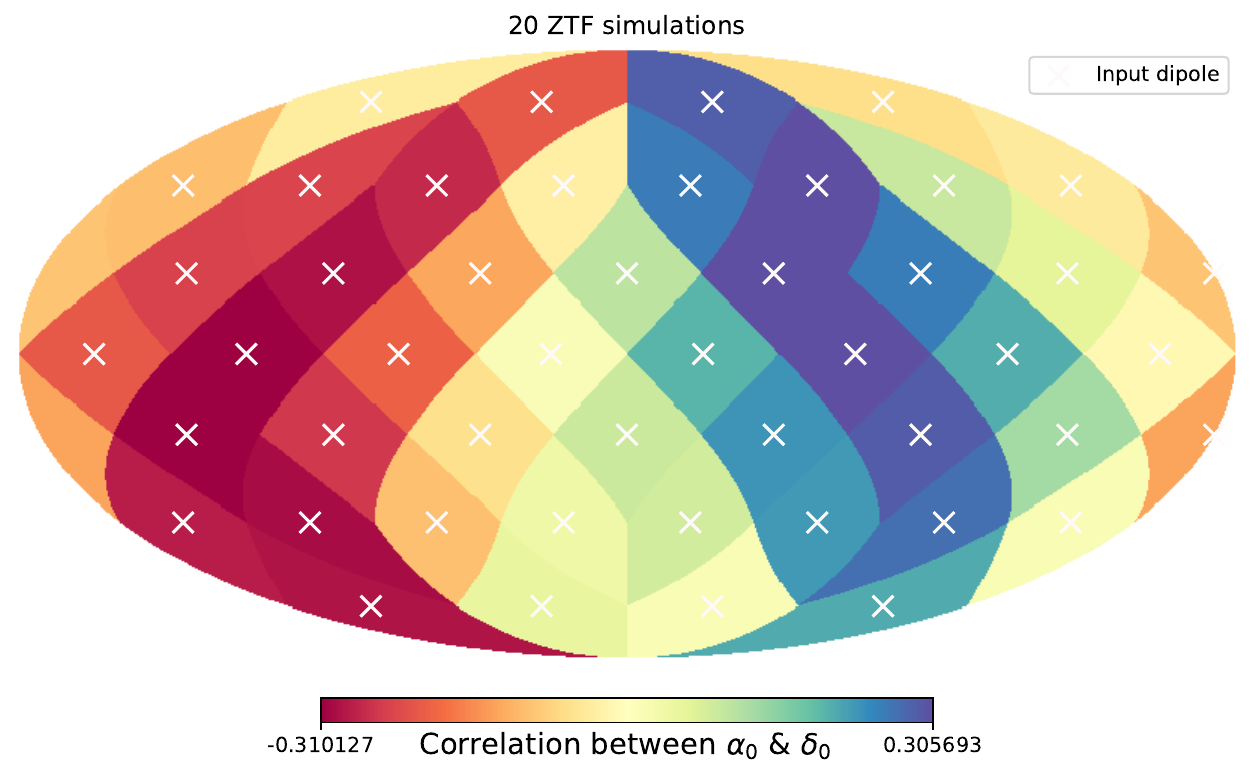}
      \caption{Skymap of the correlation between the parameter the $\alpha_0$ and $\delta_0$ of the dipole, each white cross correspond to a different location of dipole introduced in the data with the $m_B$ method, for the 20 simulations with an initial amplitude of dipole of $3~\hunit$. Each patch is colored depending of the median correlation between the two parameters for the 20 simulations.}
         \label{fig:annex_corr_ra_dec}
\end{figure}

\begin{figure}
   \centering
   \includegraphics[width=0.9\columnwidth]{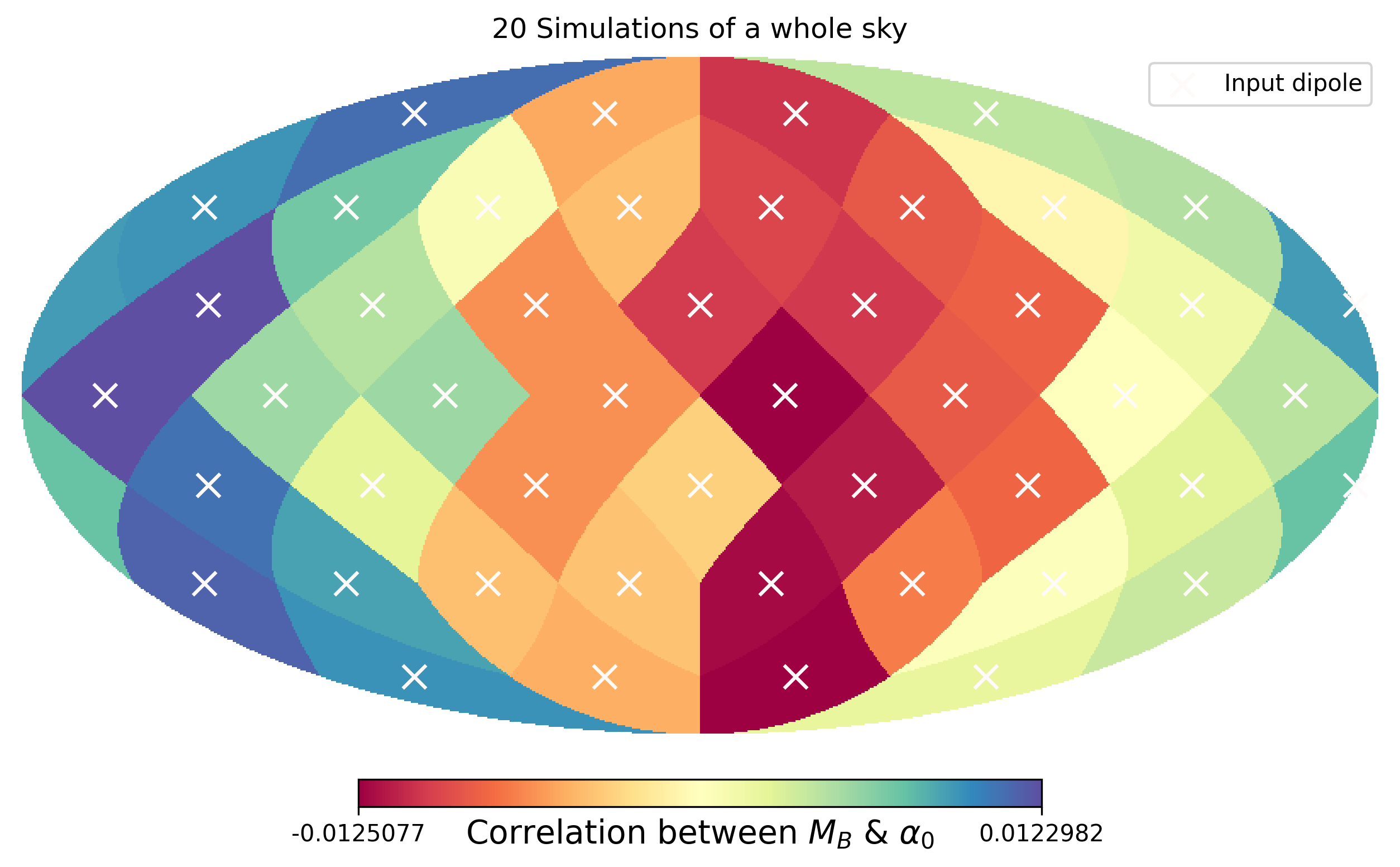}
      \caption{Skymap of the correlation between the parameter the $\alpha_0$ and $M_B$ for each dipole direction, each white cross correspond to a different location of dipole introduce in the data with the $m_B$ method, for the 20 simulations of the all sky with an initial amplitude of dipole of $3~\hunit$. Each patch is coloured in function of the median of the correlation between the two parameters for the 20 simulations.}
         \label{fig:annex_corr_Mb_ra}
\end{figure}

As explained in Sec.~\ref{sec:Dipole_analyze}, we have to fit the standardization parameters at the same time as the dipole parameters due to the correlation between $M_B$ and $H_0$ which introduced a correlation between $M_B$ and the dipole amplitude as we can see in Fig.~\ref{fig:annex_corr_Mb_H0}. We have a strong correlation in all the sky for these two parameters. However, we also have a smaller one between the $\alpha_0$ and $\delta_0$ of the dipole, as shown in Fig.~\ref{fig:annex_corr_ra_dec}. This correlation is larger near the Milky Way (MW) which also corresponds to the maximum and minimum of the oscillations discuss in Sec.~\ref{sec:discussion}. The mask on the MW makes it harder to constrain the dipole position. Fig.~\ref{fig:annex_corr_Mb_ra} shows a clear correlation between the parameters $M_B$ and $\alpha_0$ in the area of the maxima and minima of the oscillations.

\section{Results for $\alpha_0$ and $\delta_0$ for different sky coverage}
\label{appx:sky_coverage}

\begin{figure*}
   \centering
   \includegraphics[width=1.2\columnwidth]{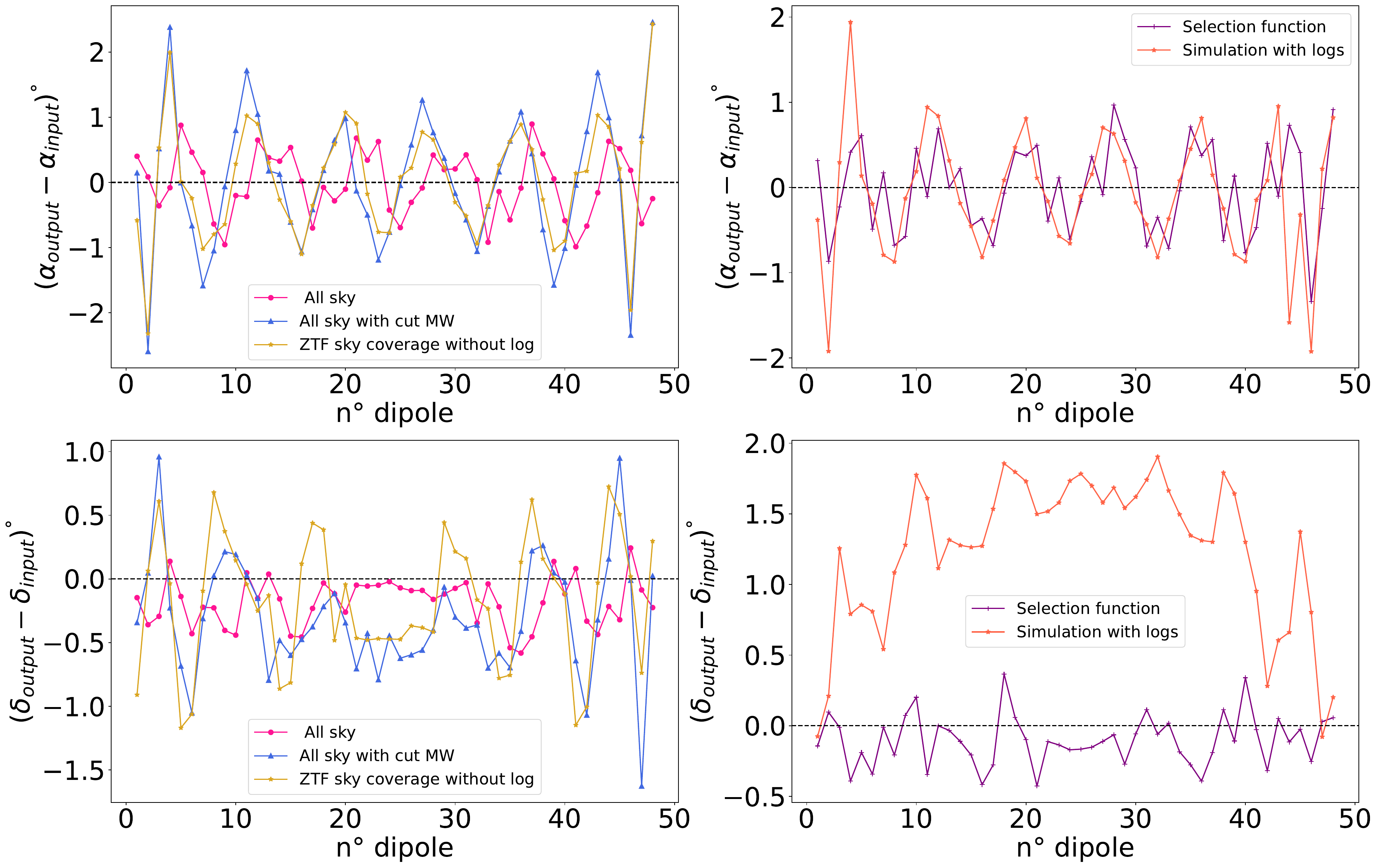}
      \caption{Comparison between three sky coverage. The results for a full sky, an uniform dispersion of SNe Ia, is represented in dark pink dot. The blue triangles correspond to simulations of the full sky but with a cut on MW. The pink stars represent simulations with a cut on MW and on declination but without logs. For all the simulation we did not use the SALT parameters but the simulated light curve parameters.}
         \label{fig:annex_sky_coverage}
\end{figure*}

\begin{figure}
   \centering
   \includegraphics[width=0.9\columnwidth]{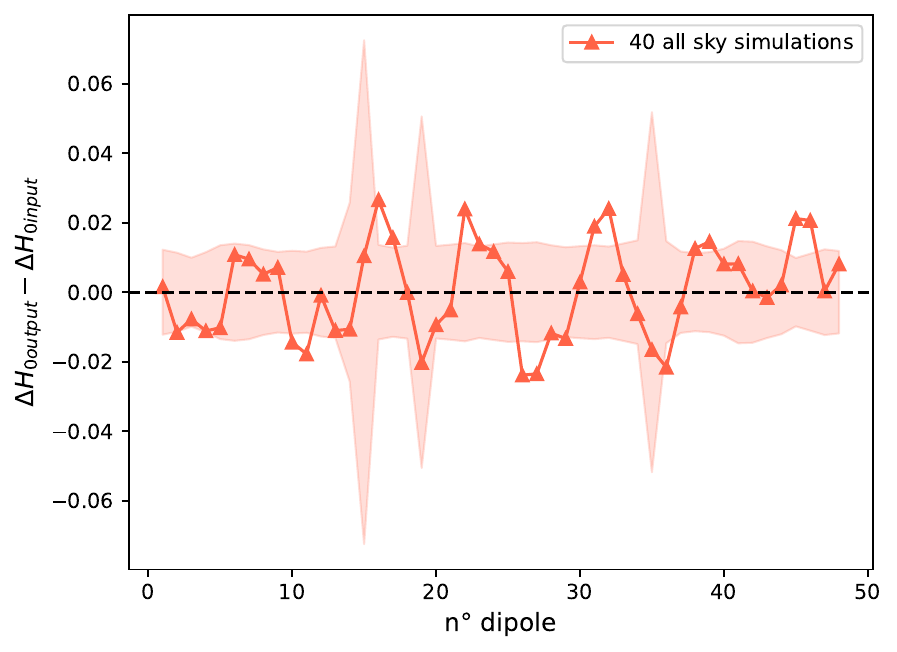}
      \caption{Difference between output and input of the amplitude of the dipole for 40 independent simulations of the full sky in function of the initial dipole introduced. With a orange band which is the error of the median (the standard deviation divided by the number of simulations).}
         \label{fig:annex_40_simu_oscillaitons}
\end{figure}

To investigate in the origin of possible bias, we do the analyses in different sky coverage in Sec.~\ref{sec:sky_coverage}, but we only present the result for the amplitude of the dipole. In this section, with the Fig.~\ref{fig:annex_sky_coverage}, we present the result of $\alpha_0$, $\delta_0$ for the five different scenarios of sky coverage. The different cases are included in the $1\sigma$ uncertainties for the dipole right ascension $\alpha_0$. In comparison, for the declination  $\delta_0$, we are biased of $\approx 1.3^\circ$ for the sky coverage with the ZTF footprint and the ZTF observing log, compare to the four other case where the bias is lower than $1\circ$. We can suppose that the difficulty in recovering the signal for the amplitude near the equatorial pole comes from the cadence of the observation and not from the lack of observations below $-30^{\circ}$. 
As presented in Appendix.~\ref{appex:correlation}, we believed that the oscillations were caused by the MW, but as seen in Fig.~\ref{fig:annex_sky_coverage} they are present for all the different sky coverage even an isotropic SN Ia distribution. However, these oscillations are included in the error as we can see for 40 simulations of the full sky in Fig.~\ref{fig:annex_40_simu_oscillaitons}. As we focus on the origin of oscillations we did 40 simulations with \texttt{skysurvey}, but we use the light curve parameters generated and we do not fit them. However, as mentioned in Sec.~\ref{sec:discussion}, these oscillations do not affect our analysis. 

\end{appendix}

\end{document}